\documentclass[reprint,prl,amsmath,amssymb,aps,superscriptaddress]{revtex4-2}
\usepackage{amssymb}
\usepackage{amsmath}

\usepackage{graphicx}
\usepackage{subfigure}

\usepackage{bm}

\usepackage{tikz}
\usetikzlibrary{matrix}
\usetikzlibrary{calc}
\newcommand{\bra}[1]{\langle #1|}
\newcommand{\ket}[1]{|#1\rangle}
\newcommand{\braket}[2]{\langle #1 | #2 \rangle}
\usepackage[sort&compress]{natbib}

\usepackage[colorlinks,
citecolor=blue
]{hyperref}

\begin{document}

	\title{Breaking the Limitations of Temporal Modulation via Mixed Continuity Conditions}

	\author{Yongge Wang}

	\affiliation{School of Physics, Harbin Institute of Technology, Harbin 150001, People’s Republic of China}
	\author{Jingfeng Yao}
	\email[]{yaojf@hit.edu.cn}

	\affiliation{School of Physics, Harbin Institute of Technology, Harbin 150001, People’s Republic of China}
	\affiliation{Heilongjiang Provincial Key Laboratory of Plasma Physics and Application Technology, Harbin 150001, People’s Republic of China}
	\affiliation{Heilongjiang Provincial Innovation Research Center for Plasma Physics and Application Technology, Harbin 150001, People’s Republic of China}
	\author{Ying Wang}

	\affiliation{School of Physics, Harbin Institute of Technology, Harbin 150001, People’s Republic of China}
	\affiliation{Heilongjiang Provincial Key Laboratory of Plasma Physics and Application Technology, Harbin 150001, People’s Republic of China}
	\affiliation{Heilongjiang Provincial Innovation Research Center for Plasma Physics and Application Technology, Harbin 150001, People’s Republic of China}
	\author{Chengxun Yuan}
	\email[]{yuancx@hit.edu.cn}

	\affiliation{School of Physics, Harbin Institute of Technology, Harbin 150001, People’s Republic of China}
	\affiliation{Heilongjiang Provincial Key Laboratory of Plasma Physics and Application Technology, Harbin 150001, People’s Republic of China}
	\affiliation{Heilongjiang Provincial Innovation Research Center for Plasma Physics and Application Technology, Harbin 150001, People’s Republic of China}
	\author{Zhongxiang Zhou}
	\email[]{zhouzx@hit.edu.cn}

	\affiliation{School of Physics, Harbin Institute of Technology, Harbin 150001, People’s Republic of China}
	\affiliation{Heilongjiang Provincial Key Laboratory of Plasma Physics and Application Technology, Harbin 150001, People’s Republic of China}
	\affiliation{Heilongjiang Provincial Innovation Research Center for Plasma Physics and Application Technology, Harbin 150001, People’s Republic of China}

	\date{\today}
	
	\begin{abstract}
The conventional description of time-varying media assumes that electromagnetic fields evolve according to fixed continuity conditions during parameter jumps. Here we reveal that these conditions are not physical constraints but tunable design degrees of freedom. By developing a unified framework that treats continuity rules as engineerable parameters,  we expand the scope of time-varying metamaterials and enable wave phenomena previously considered impossible. For instance, non-resonant, reflectionless wave amplification without momentum bandgaps, and reversible conversion between propagating waves and static fields for optical memory, etc. This work opens a new dimension for controlling light-matter interactions.

	\end{abstract}
	
	\maketitle

Time-varying electromagnetic media, whose material parameters are modulated in time, have attracted growing interest in recent years. Time modulation can exhibit numerous intriguing optical properties and give rise to a variety of exotic phenomena absent in static media, stationary charge radiation\cite{li2023stationary}, Faraday rotation\cite{PhysRevLett.128.173901,he2023faraday}, Wood anomalies\cite{PhysRevLett.125.127403}, and anomalous Cherenkov radiation\cite{doi:10.1073/pnas.2119705119}. The most significant phenomenon among them are time reflections and time refractions occurring at temporal interfaces\cite{PhysRevLett.132.263802,WOS:001575855000001,WOS:001604755900027,WOS:001287600900019,PhysRevLett.133.186902, WOS:000531425900022,WOS:001258302000023,doi:10.1126/sciadv.adz5445}. When an incident wave encounters a sudden change in material parameters, it typically splits into two counter-propagating waves. Under periodic or disordered modulation, these waves interfere and can form momentum band gaps, leading to the concept of photonic time crystals\cite{PhysRevLett.126.163902,doi:10.1073/pnas.2119705119,PhysRevLett.128.094503,doi:10.1126/science.abo3324,PhysRevA.79.053821,Lustig:18,Sharabi:22}. Modes inside these gapsmay grow exponentially by drawing energy from the modulation, offering a mechanism for wave amplification\cite{PhysRevLett.126.163902,doi:10.1073/pnas.2119705119,PhysRevLett.128.094503,doi:10.1126/science.abo3324}.

Despite experimental challenges, a number of studies have successfully observed time refraction, time reflection, and momentum band gaps\cite{PhysRevLett.128.094503,10.1063/1.4928659,WOS:001604755900027,WOS:001287600900019,doi:10.1126/sciadv.adg7541,doi:10.1126/science.aah6822}. Research has also extended to dispersive materials, which are more general and physically realistic\cite{mirmoosa2022dipole,PhysRevB.103.144303,PhysRevLett.134.183801}. Yet in modeling these time-varying dispersive media, the continuity conditions imposed on electromagnetic fields at a temporal jump are typically assumed without question. In reality, how these fields evolve at an interface is determined by the specific physical process that modulates the material\cite{ WOS:001575855000001,GaliffiYinAl}. In Newton's initial derivation of the speed of sound, the relationship between gas pressure and volume was described using Boyle's law, which was the only available gas law at the time\cite{newton1833philosophiae,10.1119/1.1970365}. This led to a discrepancy between his theoretical prediction and experimental measurements\cite{laplace1816vitesse,pierce2019acoustics,lighthill2001waves}. Similarly, in the context of temporal modulation, the relations governing the field quantities and the modulating parameters should not be implicitly assumed; otherwise, the resulting theory may fail to capture the underlying physical reality.
In this work, we develop a general theoretical framework that incorporates the effects of different continuity conditions. Furthermore, we demonstrate that these mixed conditions are not merely a modeling refinement, but a pathway to new physical phenomena. Our work thus introduces a new degree of freedom to the study of time-varying electromagnetic media, paving the way for future experimental explorations and applications.

We begin with the electromagnetic field equations for static dispersive media and subsequently generalize them to the temporally varying case.
	\begin{equation}
	\begin{aligned}
		\nabla \times\bm H=\varepsilon _0 \varepsilon_\infty \frac{\partial \bm E}{\partial t}+\bm J_\text{Source}+\bm J_\text{p}\label{eq1}.
	\end{aligned}
\end{equation}
Here, $\bm H$ is the magnetic field, $\bm E$ is the electric field, $t$ is time, and $\varepsilon_\infty$ is the background permittivity. $\bm J_\text{Source}$ is the applied current source, and $\bm  J_\text p$ is the polarization current density, which characterizes the material dispersion and satisfies $\bm J_\text{p}=\dot {\bm P}$, where the dot denotes the time derivative. We can further write the equation for polarization $\bm P$ as

\begin{equation}
	\begin{aligned}
		\frac{\partial^2\bm P}{\partial t^2}+\nu\frac{\partial\bm P}{\partial t} +\omega_0^2\bm P=\varepsilon_0 \omega_{p}^2\bm E(t).\label{eq2}
	\end{aligned}
\end{equation}
$m_e$ denotes the electron mass, $\nu$ represents the collision frequency, $\omega_0$ is the resonance frequency and $\omega_{p}$ is the plasma frequency. 

For temporally varying material parameters, the electric field can be expressed as $\bm E(\bm x ,t;\bm u)$
, where $\bm x$ is the spatial coordinate and $ \bm u =\{u_1,u_2,\dots\}$ collectively represents the set of time-varying parameters (e.g., $\varepsilon_\infty$ in Eq.~\ref{eq1} and $\omega_0$ in Eq.~\ref{eq2}). The total time derivative of the electric field, accounting for both field variations and parametric changes, can then be written as:

\begin{equation}
	\begin{aligned}
		\dot{\bm E}=\frac{\mathrm d \bm E}{\mathrm dt}=\sum_i  \frac{\partial \bm E}{\partial u_i}\dot{u}_i+\frac{\partial\bm E}{\partial t}\label{eq3}.
	\end{aligned}
\end{equation}The operator $\partial /\partial u_i$ denotes the partial derivative with respect to $u_i$, while keeping $t$ and  all other parameters $u_j$ (with $i\neq j$) fixed, i.e.,
\begin{equation}
	\begin{aligned}
	\frac{\partial}{\partial u_i}\stackrel{\text{def}}{=}\left(\frac{\partial}{\partial u_i}\right)_{t,\{u_j\}_{j\neq i}}.
	\end{aligned}
\end{equation}
This differential describes a polytropic process, which we define here as a scenario where no internal energy exchange occurs within the wave system, and all energy variations arise solely from external modulation.
Substituting Eq.~\ref{eq3} into Eq.~\ref{eq1} gives the evolution equation for the electric field under temporal parameter variation:
\begin{equation}
\begin{aligned}
	\varepsilon_0\dot{\bm E}=\frac{1}{\varepsilon_\infty }\left(\nabla \times \bm H-\bm J_{\text{Source}}-\bm J_{\text{p}}\right)+\varepsilon _0 \sum_i \frac{\partial \bm E}{\partial u_i} \dot u_i.\label{eq5}
\end{aligned}
\end{equation}

In the dispersion-free case where $\bm J_\mathrm p=0$, and assuming $(\varepsilon_\infty) ^\gamma\bm E$ remains temporally continuous during a permittivity jump, we obtain:

\begin{equation}
\begin{aligned}
	 \sum_i \frac{\partial \bm E}{\partial u_i} \dot u_i=\frac{\partial \bm E}{\partial\varepsilon_\infty}\frac{\mathrm d \varepsilon_\infty }{\mathrm d t}=-\gamma\frac{ \bm E}{\varepsilon_\infty}\frac{\mathrm d \varepsilon_\infty }{\mathrm d t}.\label{eq6}
\end{aligned}
\end{equation}
Combining Eq.~\ref{eq6} with Eq.~\ref{eq5} recovers the conventional form of the field equations widely used to describe electromagnetic behavior under permittivity modulation.
It can be seen that $\gamma=0$ corresponds to the continuity of the electric field, whereas $\gamma=1$ corresponds to the continuity of the electric displacement field—the scenario most extensively studied in conventional works. Here, $\gamma$ is referred to as the polytropic index of the process. At $\gamma=1/2$, the system becomes adiabatic, and no energy growth of waves can be drawn from the modulation (see Supplementary Material).
 More generally, $\partial \bm E/\partial u_i$ is governed by the microscopic details of the modulation process. 

\begin{figure}[b]
	
	\includegraphics[width=0.8\linewidth]{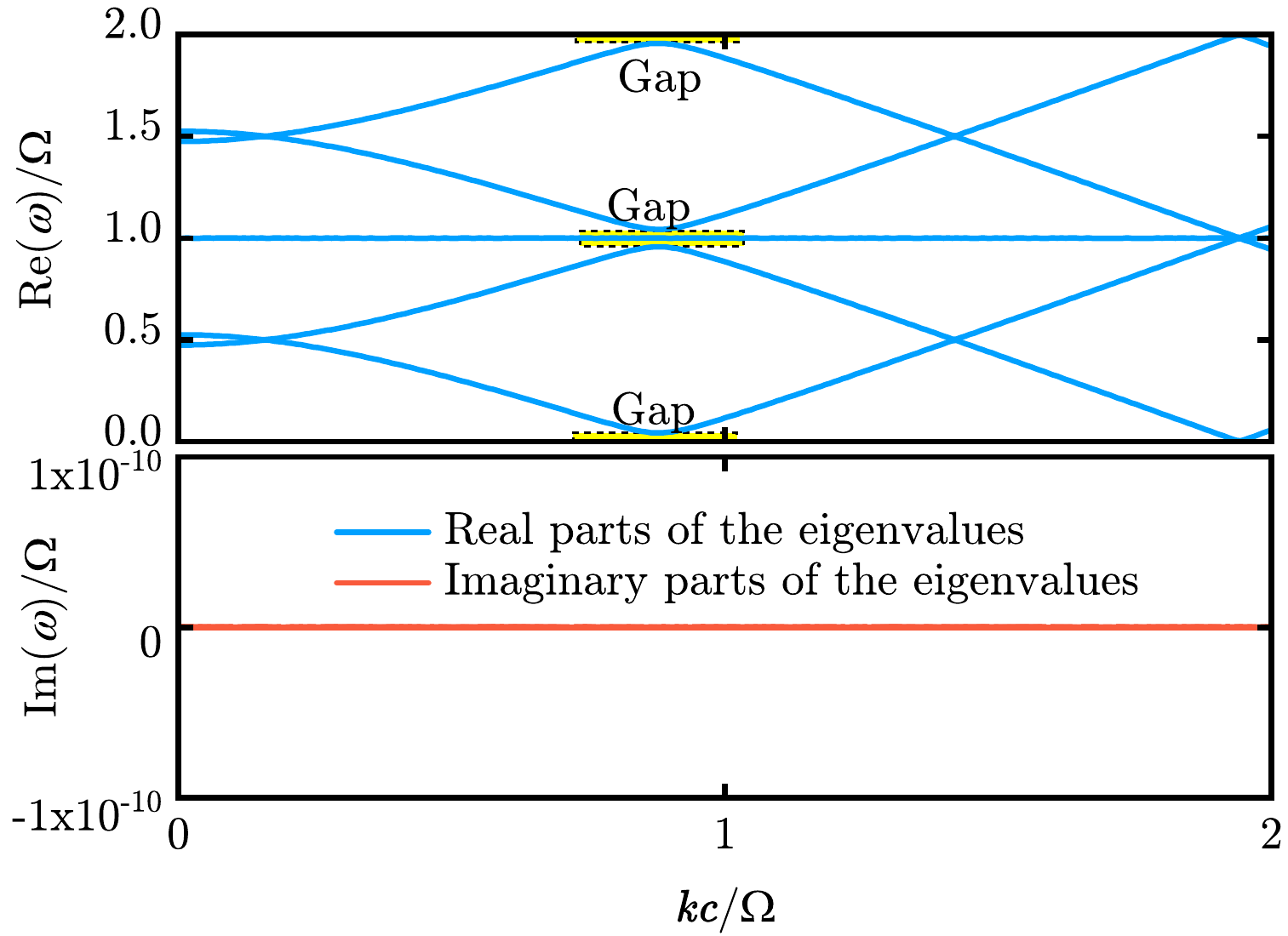}
	
	\caption{\label{fig0} Real and imaginary parts of the time-modulated Drude dispersive medium for $J_p\omega_{p}^{-1}=\text{const.}$ continuity condition, with band gaps indicated.}
\end{figure}

When considering dispersive materials, we can express the polarization field as $\bm P(\bm x ,t;\bm u)$, with its time derivative given by
$
		\bm J_\text{p}=\dot{\bm P}=\sum_i  \frac{\partial \bm P}{\partial u_i}\dot{u}_i+\frac{\partial\bm P}{\partial t}.
$
 Similarly, current density has the form ${\bm J}_{\text{p}}(\bm x ,t;\bm u,\dot{\bm u})$, whose derivative is	
$\dot { {\bm J}}_\text{p}=\sum_i  \left(\frac{\partial  {\bm J}_\text{p}}{\partial u_i}\dot{u}_i+\frac{\partial \bm J_{p}}{\partial\dot u_i}\ddot u_i +\frac{\partial ^2\bm P}{\partial t\partial u_i}\dot u_i\right)+\frac{\partial ^2 {\bm P}}{\partial t^2}.
$

Thus, we can express the evolution of $ {\bm J}_\text p$ in the time-varying form as
\begin{equation}
	\begin{aligned}
		\dot { {\bm J}}_\text{p}-\sum_i\left(\frac{\partial  {\bm J}_\text{p}}{\partial u_i}\dot{u}_i+\frac{\partial \bm J_{p}}{\partial\dot u_i}\ddot u_i +\frac{\partial ^2\bm P}{\partial t\partial u_i}\dot u_i\right)\\
		+\nu(t)\left(\dot{\bm P}-\sum_i  \frac{\partial \bm P}{\partial u_i}\dot{u}_i\right)
		+\omega_0^2(t)\bm P=\varepsilon_0 \omega_{p}^2(t) \bm E(t).\label{eq11}
	\end{aligned}
\end{equation}

In a Drude-type dispersive medium, the energy density stored in the form of current density reads $E_J=\frac 12  |J_p|^2/(\varepsilon_0\omega_{p}^2)$. Hence, if $J_p/\omega_{p}$ is assumed to be continuous in time, no energy can be drawn from the system, and consequently no energy band gap opens under temporal modulation. In Fig.~\ref{fig0}, we plot the band structure for plasma frequency modulation under this continuity condition, with $\omega_{p}^2(t)=0.23\,\Omega^2(1+0.4\cos(\Omega t))$ as shown. Here $\Omega$ is the modulation frequency. It is observed that, even in the presence of modulation, neither a momentum band gap nor an imaginary part of the eigenvalues appears. In the yellow-shaded region of Fig.~\ref{fig0}, however, an energy band gap is opened by the temporal modulation. Such a gap is not typically observed in the conventional time photonic crystal; instead, it resembles a 0-gap in Floquet systems. It should be emphasized that this 0-gap does not arise from a time-averaged Hamiltonian (the time-varying part averages to zero), but rather from a resonance-induced band anticrossing.

\begin{figure}[htb]
\vspace{1em}
\subfigure[]{\parbox[][0.35\linewidth][c]{0.54\linewidth}{\includegraphics[width=\linewidth]{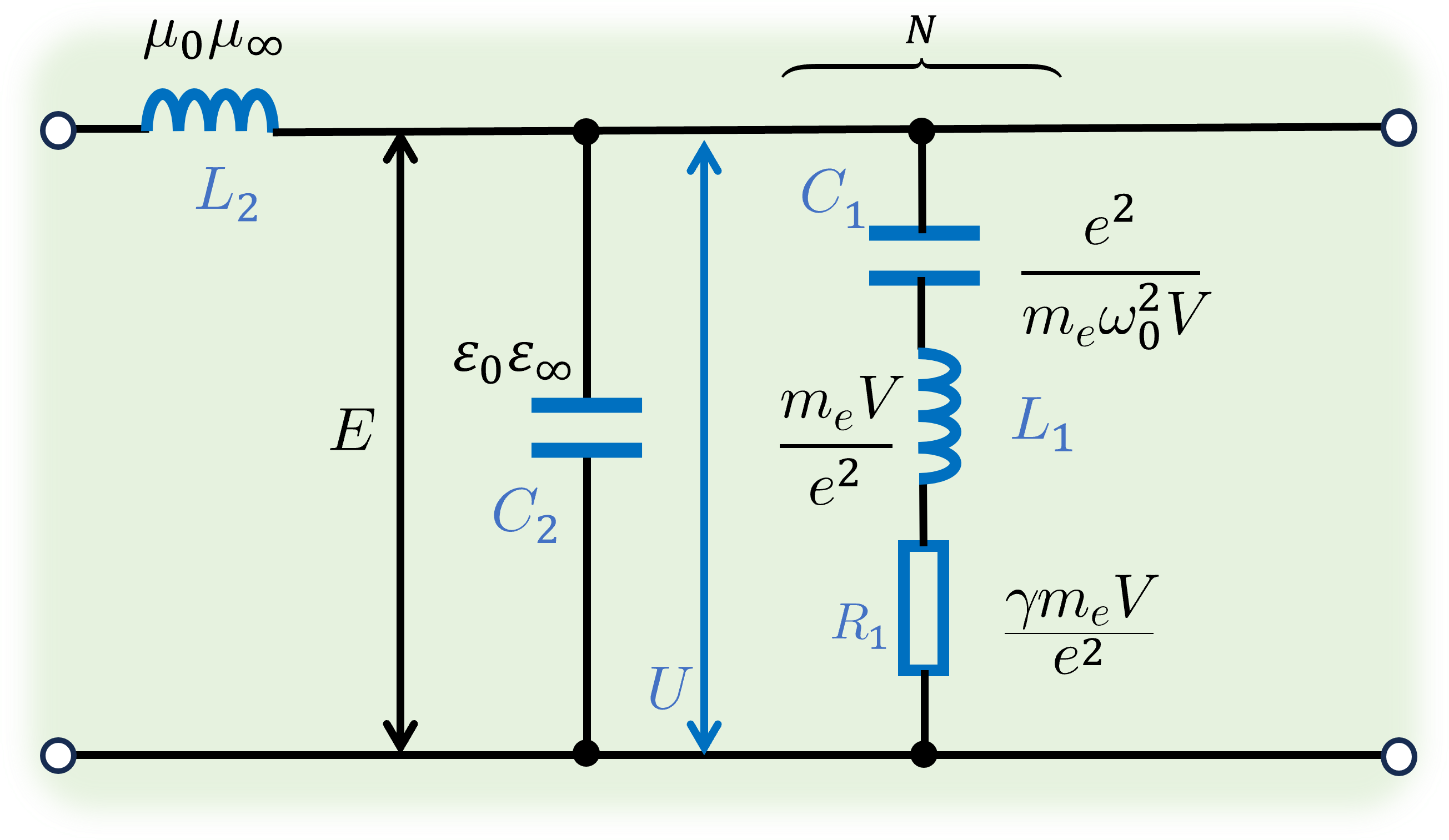}}}
\hspace{0.4em}
\subfigure[]{\parbox[][0.35\linewidth][c]{0.185\linewidth}{\includegraphics[width=\linewidth]{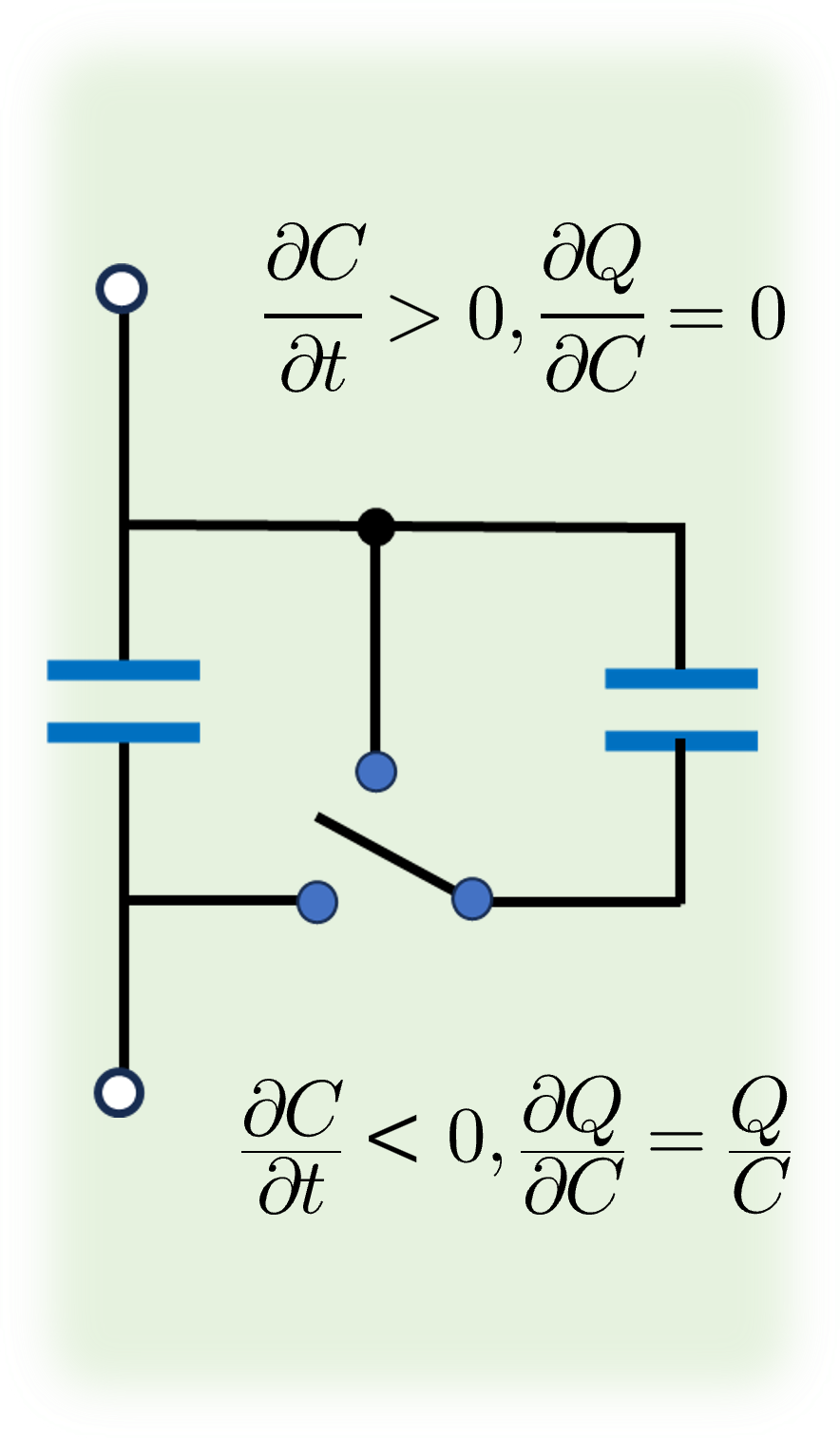}}}
\hspace{0.5em}
\subfigure[]{\parbox[][0.35\linewidth][c]{0.185\linewidth}{\includegraphics[width=\linewidth]{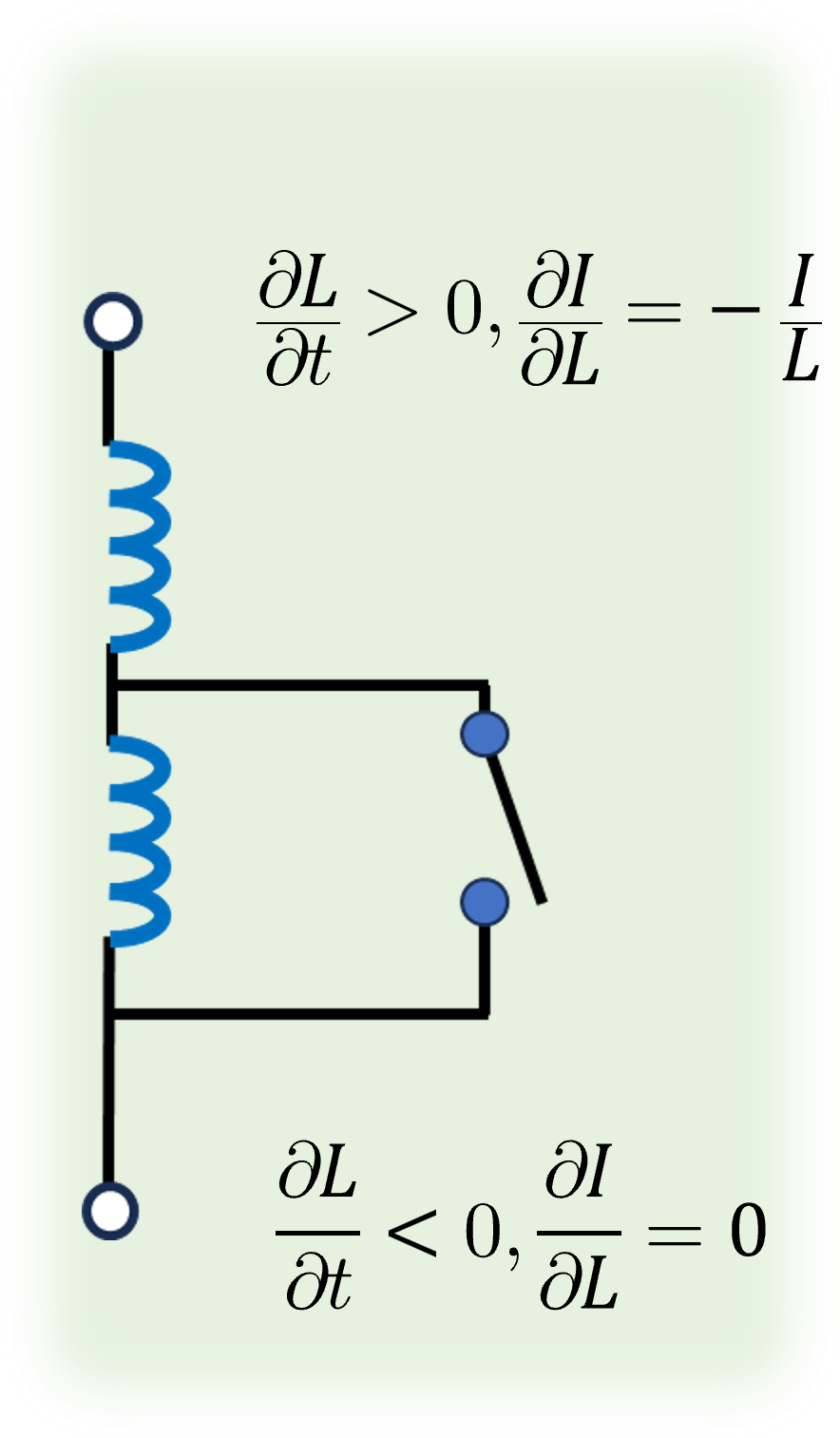}}}
\subfigure[]{\parbox[][0.35\linewidth][c]{0.55\linewidth}{\label{fig6a}		\includegraphics[width=\linewidth]{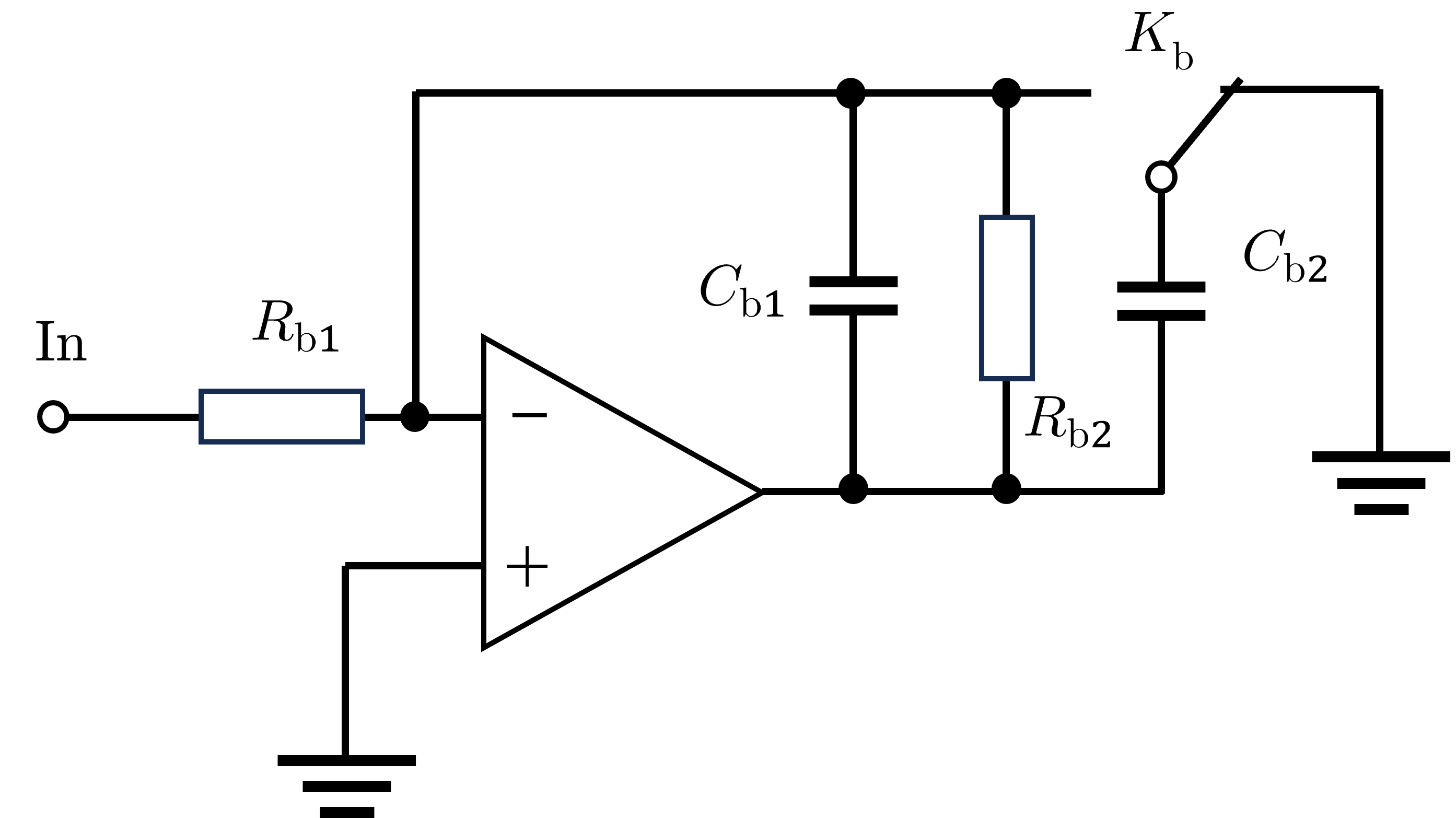}}}
\hspace{1.2em}
\subfigure[]{\parbox[][0.35\linewidth][c]{0.38\linewidth}{\label{fig6b}		\includegraphics[width=\linewidth]{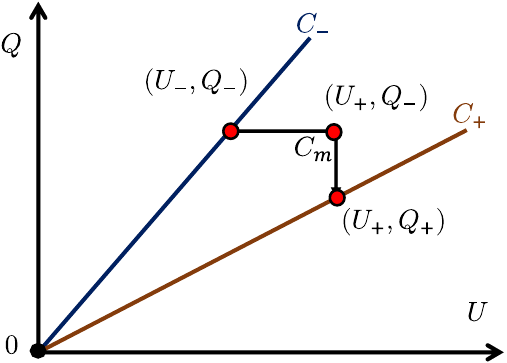}}}
\caption{\label{fig1} Analogical model for time-modulated material dispersion response. (a) Transmission line model. (b) Passive modulation of capacitance. (c) Passive modulation of inductance. (d) Practically realizable capacitance modulation circuit based on an operational amplifier. Adjusting the ratio of $R_{\mathrm b 2}$ to $R_{\mathrm b1}$ tunes the equivalent capacitance while keeping the charge constant. Meanwhile, toggling switch $K_\mathrm b$ connects $C_{\mathrm b1}$ and  $C_{\mathrm b 2}$ in parallel, enabling a voltage-conserving capacitance change, and vice versa. (e) Schematic representation of the decomposition of a capacitance jump from $C_-\rightarrow C_+$ into a charge-conserving step and a voltage-conserving step.}
\end{figure}

In our method, the additional terms describe different continuity conditions for the electromagnetic parameters, offering a new degree of freedom that allows for novel physical effects. To intuitively illustrate these continuity conditions, we develop a lumped-element circuit analogy, as illustrated in Fig.~\ref{fig1}.

\begin{figure*}[!t]
	
	\subfigure[\label{fig2a}]{\includegraphics[width=0.22\linewidth]{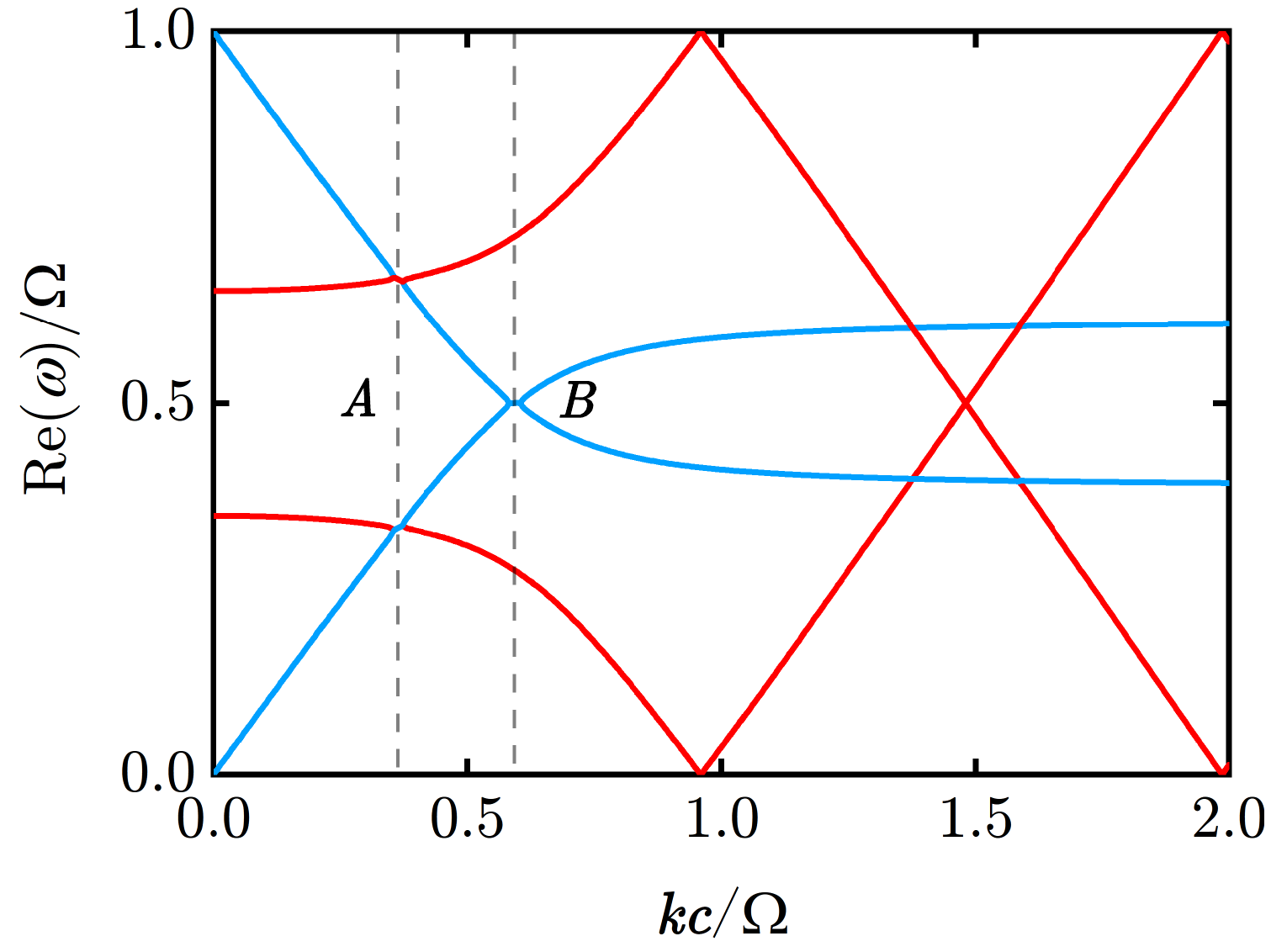}}
	\subfigure[\label{fig2b}]{\includegraphics[width=0.225\linewidth]{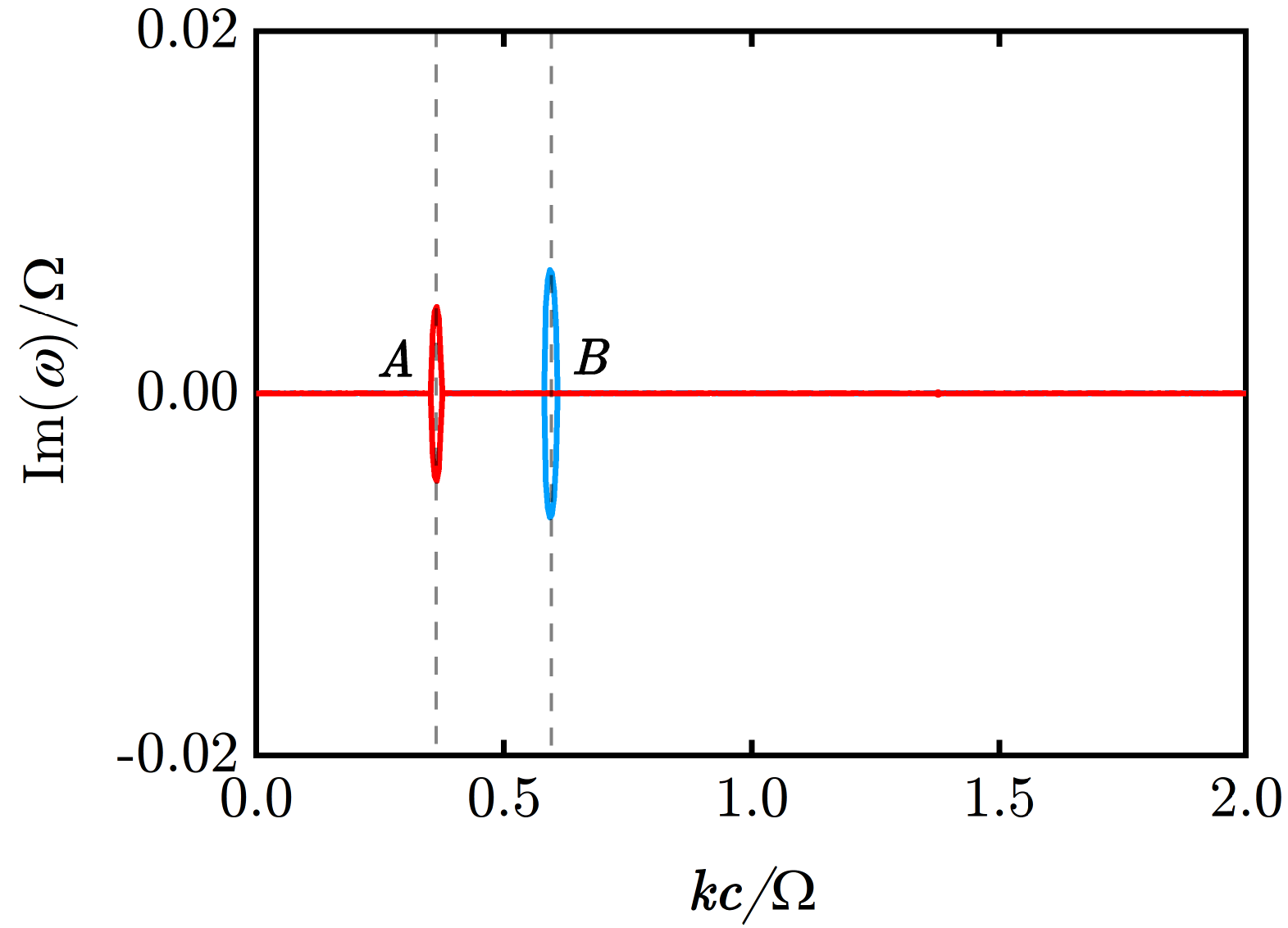}}
	\subfigure[\label{fig2c}]{\includegraphics[width=0.22\linewidth]{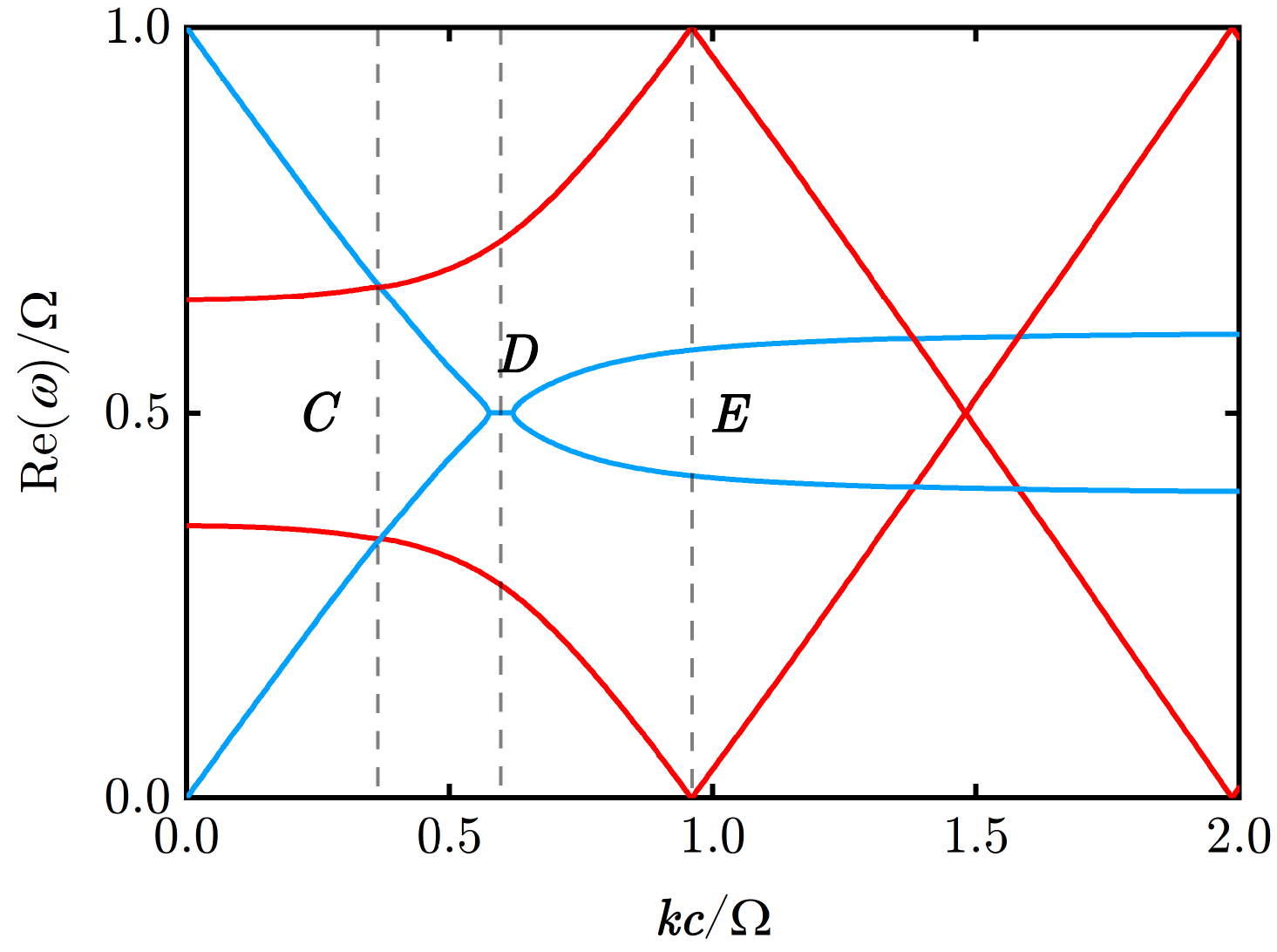}}
	\subfigure[\label{fig2d}]{\includegraphics[width=0.225\linewidth]{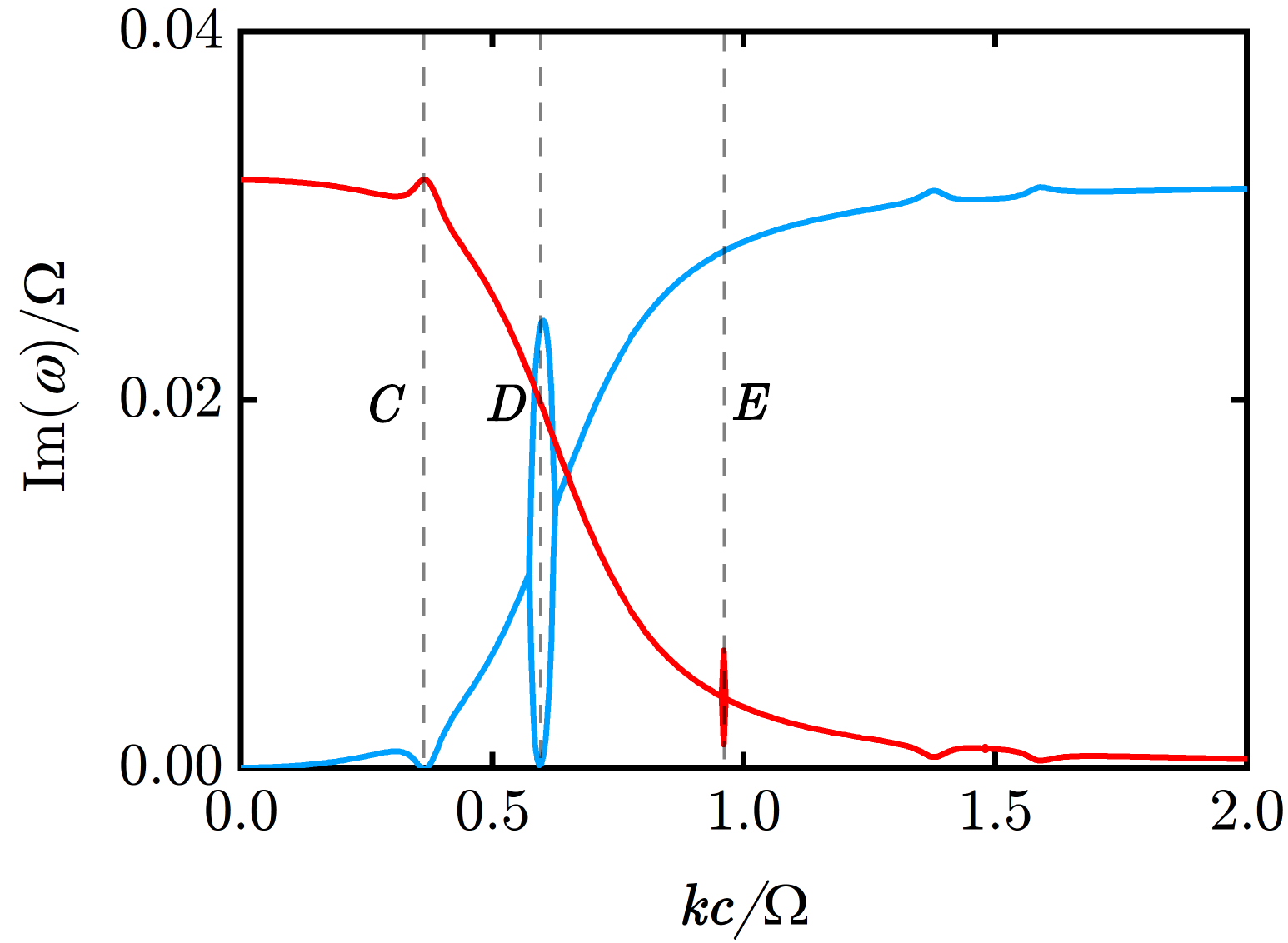}}
	\caption{\label{fig2} Band structure under $\omega_0$ modulation. The real parts (a) and imaginary part (b) of the energy band under traditional modulation approach, i.e., $\partial \bm P/\partial \omega_0^2 =0$. The real parts (c) and imaginary part (d) of the energy band under passive modulation approach.}
\end{figure*}

Fig.~\ref{fig1}(a) illustrates a transmission line model where the capacitance $C_2$ of the circuit corresponds to background permittivity $\varepsilon_0\varepsilon_\infty$, and the voltage $U$ across its terminals corresponds to $E$. On the right side, $N$ series $RLC$ resonant circuits are connected in parallel, with the values of $R_1$, $L_1$ and $C_1$ corresponding to $\nu m_eV/e^2$, $m_eV/e^2$, and $e^2/(m_e\omega_0^2V)$ respectively. Here, $N$ is the number of oscillators per unit volume $V$. Figs.~\ref{fig1}(b) and (c) illustrate passive modulation schemes for capacitors and inductors. The modulation of a time-varying capacitor $C(t)$ can proceed under a constant terminal voltage $U$, a constant charge $Q$, or an intermediate condition between these two extremes. For a time-varying inductor $L(t)$, modulation can preserve a constant current $I$, a constant flux linkage (i.e., the product of current and inductance), or a state between these two.
In the passive modulation case, reversing the modulation direction gives rise to different continuity conditions. Consequently, the electromagnetic wave cannot draw energy from the modulation and, consequently, exhibits pure attenuation.

To illustrate the idea of mixed continuity, we introduce a physically realizable circuit model, which provides a platform for constructing temporally modulated metamaterials and metasurfaces. Depicted in Fig.~\ref{fig1}(d) is an operational-amplifier-based capacitance-multiplying circuit. When $R_{\mathrm b2}\ll R_{\mathrm b1}$ , the input impedance is equivalent to a capacitance $C_{\mathrm 1}R_{\mathrm  b 2}/R_{\mathrm b1}$. The magnitude of this equivalent capacitance can be modified by varying $R_{\mathrm b1}$ and $R_{\mathrm 2}$while the charge on it remains invariant during the tuning process. Meanwhile, capacitor $C_{\mathrm b2}$, being grounded at one terminal, draws current from the source so as to maintain a voltage identical to that across $C_{\mathrm b1}$. Consequently, upon toggling the switch $K$, the equivalent capacitance changes while the voltage across it remains constant. When the capacitance undergoes a jump from $C_-$ to $C_+$ with the constraint $U_+/U_-=(Q_+C_-)/(Q_-C_+)$, the process can be decomposed into two distinct capacitance modulation stages: a charge-conserving transition from $C_-$ to 
intermediate state $C_\mathrm m$, followed by a voltage-conserving transition from $C_\mathrm m$ to $C_+$, as depicted in Fig.~\ref{fig6b}.

As a specific example, we consider the modulation of $\omega_0^2$, assuming $\bm P$ to be continuous for a downward jump and $\omega_0^2  P$ continuous for an upward jump. That is to say, $\partial \bm P/\partial \omega_0^2=-\theta(\dot{\omega_0^2})\bm P / \omega_0^2$. Here, $\theta $ is the step function. And we further have
\begin{equation}
	\begin{aligned}
		\frac{\partial^2 \bm P}{\partial t\omega_0^2}
		&=-\theta(\dot{\omega_0^2})\frac{1}{\omega_0^2}\left(\dot{\bm P}-\frac{ \bm P}{ \omega_0^2}\dot{\omega_0^2}\right).\label{e8}
	\end{aligned}
\end{equation}

The coefficient of the $\dot {\bm P}$ term appearing here represents the modulation-induced energy dissipation.

In our model, we use the modulation frequency $\Omega$ as the normalization frequency unit, and we set the square of the plasma frequency to $\omega_\mathrm p^2 =0.05\,\Omega^2$. The modulation of the resonance frequency is expressed as $\omega_0^2(t)=0.375\,\Omega^2(1+\Delta \cos(\Omega t))$, where $\Delta$ is the modulation depth and is set to $\Delta=0.1$.

We first calculate the band structure under the assumption of continuous 
 $\bm P$ (i.e. $\partial \mathbf{P} / \partial \omega_0^2 = 0$) in Fig.~\ref{fig2a} and \ref{fig2b}, where $\partial \bm P /\partial \omega_0^2 =0$. Here, $k$ is the wave number. 
A harmonic time dependence $\exp(\mathrm i \omega t)$ is adopted. The blue curve in the Fig.~\ref{fig2} corresponds to the first energy band under static dispersion, while the red curve corresponds to the second band. In this conventional scheme, the modulation can induce exponential growth of the modes. Two momentum gaps can be observed in the figure, appearing at two distinct momentum positions, $A$ and $B$. 

As a comparison, the band structure under the proposed passive modulation is plotted in Figs.~\ref{fig2c} and \ref{fig2d}. Under such time modulation, the imaginary part of the band is observed to split at low $k$. 
In fact, this implies that the passive modulation scheme introduces an effective static loss analogous to $\nu$. However, this introduces an energy dissipation associated with the electric polarization field $\bm P$ rather than the current density field $\bm J_\text{p}$. This originates from the zero-frequency component of the coefficient of $\dot {\bm P}$ in Eq.~\ref{e8}. The time derivative term becomes significant at higher modulation frequencies, which extends beyond the conventional temporal modulation framework. 

At low $k$, the first band deviates further from the resonance frequency, so the energy is primarily stored in the electric field. In contrast, the second band lies closer to the resonance frequency, where energy is mainly transferred between the polarization field and the current density field, thereby resulting in greater loss. 

At point $D$, the intersection of the same first band at different orders results in a behavior analogous to a conventional momentum band gap. However, the difference is that a pair of modes degenerate in their real parts exhibit different attenuation rates. 
One experiences strong attenuation, while the other exhibits almost no loss. This distinction can lead to mode selection and time-reflection phenomena. At point $E$, an additional momentum bandgap appears at $\omega=\Omega$ compares to Figs.~\ref{fig2a} and \ref{fig2b}. 
In our approach, however, the asymmetry between the positive and negative half cycles of the modulation introduces an infinite series of higher-order harmonics via the $\theta$ function. A first-order perturbation thus also participates at the frequency $\Omega$, giving rise to an additional band gap.

Having established the band structure modifications, we now investigate the dynamical consequences of a single parameter jump. 
In a static medium, a plane wave solution can be expressed as a superposition of four basis vectors ${\ket{e},\ket{h},\ket{j},\ket{p}}$ corresponding to the electric field, magnetic field, current density, and polarization density. the time evolution of a linearly polarized plane wave can be represented by four complex coefficients $(\tilde{E},\tilde{H},\tilde{J},\tilde{P})^{\mathsf T}$ in the basis of {${\ket{e},\ket{h},\ket{j},\ket{p}}$}:
\begin{equation}
	\ket{\psi(t)}=(\tilde{E}(t)\ket{e}+\tilde{H}(t)\ket{h}+\tilde{J}(t)\ket{j}+\tilde{P}(t)\ket{p}),
\end{equation}
follow the time evolution equation
\begin{equation}
-\mathrm i\frac{\mathrm d}{\mathrm d t}\begin{pmatrix}  \tilde E \\\tilde H \\\tilde J \\\tilde P \end{pmatrix}	=\hat{M}(k)\begin{pmatrix}\tilde E \\\tilde H \\\tilde J \\\tilde P  \end{pmatrix}.
\end{equation}
Considering the polarization with the magnetic field along the $x$-direction and the electric field along the 
$z$-direction, the wave propagating in the $y$-direction can be expressed as
$\ket{\psi(t)}\mathrm e^{-\mathrm ik_{y}x}$. Within the representation of $\ket{e,h,j,p}$, the eigenmodes are determined by solving the following eigenvalue equation:
$
	\hat M(k)\ket{\phi_i}=\omega_i \ket{\phi_i},
$
where $\omega_i$ is the eigenfrequency and $\ket{\phi_i}$ is the correspondent eigenstate. Here, $\hat M$ is
\begin{equation}
	\hat M(k)=
\begin{tikzpicture}[baseline=(current bounding box.center)]
	\matrix (m) [matrix of math nodes, left delimiter=(, right delimiter=)]
	{
		0 & \frac{k_y}{\varepsilon_0\varepsilon_\infty}  & \frac{\mathrm i}{\varepsilon_0\varepsilon_\infty}& 0\\
		\frac{k_y}{\mu_0\mu_r} & 0 & 0 & 0\\		
		-\mathrm i \varepsilon_0\omega_\mathrm p^2 & 0& 0 & \mathrm i \omega_0^2 \\
		0 & 0  & -\mathrm i& 0\\
	};
	\draw[dashed,red, very thick] ($(m-1-1.north west)+ (-6pt, 2pt)$) rectangle ($(m-2-2.south east) + (5pt,-4pt)$);
	\draw[dashed,blue, very thick] ($(m-1-1.north west)+ (-8pt, 4pt)$) rectangle ($(m-3-3.south east) + (3pt,0pt)$);
	\draw[dashed,green, very thick] ($(m-1-1.north west)+ (-10pt, 6pt)$) rectangle ($(m-4-4.south east)+(2pt,-0pt)$);
\end{tikzpicture}
.\label{eq21}
\end{equation}

Here, the red, blue, and green boxes in Eq.~\ref{eq21} correspond to the cases of linear dispersion, Drude dispersion, and Lorentz dispersion, respectively, without considering loss. Therefore, they possess two, three, and four eigenmodes, respectively. This matrix can be further extended when more Lorentz oscillators are considered, where each additional oscillator introduces two components, $J_i$ and $P_i$, and yields a pair of positive- and negative-frequency modes.

Note that due to the non-Hermiticity of 
$\hat M$, the eigenstates are not orthonormal under the standard inner product. We define a metric $G=(g_{i,j})$ to further refine the projection of the modes. It can be expressed via the inner product of $\mathbb{C}^4$:
$
	g_{ij}=\braket{\phi_i}{\phi_j}.
$
Consequently, we obtain the dual basis as

$
	\ket{\phi^i}=\sum_j {g^{ji}} \ket{\phi_j}.
$
Where $g^{ij}$ are the elements of the inverse metric, satisfying  $G^{-1}=(g^{ij})$. The temporal evolution of any given state is determined by its initial state.

$
	\ket{\psi(t)}=\sum_i  \mathrm e^{\mathrm i \omega t} c_i\ket{\phi_i}=\sum_i \mathrm e^{\mathrm i \omega t} \braket{\phi^i}{\psi(0)} \ket{\phi_i}.
$
Here, $c_i$ is the expand coefficient of $\ket{\phi_i}$.

Following a parameter jump, the field quantities are scaled according to the different continuity conditions as aforementioned, and can thus be described by a transformation operator:
$
	\hat T=\sum _{i=e,h,j,p}\ket{i}\alpha_i\bra{i}.
$
Here, $\alpha_i$ represents the scaling factor for the field labeled by $i$ (e.g., $\alpha_e$ for the electric field). Hence, after the sudden change in time (from parameter 1 to parameter 2), the state can be expressed in the new basis as
\begin{equation}
	\begin{aligned}
		\ket{\psi_2}&=\sum_i c_{2,i}\ket{\phi_{2,i}}=\sum_i \bra{\phi^{2,i}}\hat{T}_{1\rightarrow 2}\ket{\psi_1} \ket{\phi_{2,i}}.
		\end{aligned}
\end{equation}
The coefficients for state transitions are thus expressed as $
		c_{2,j}/c_{1,i}=\bra{\phi^{2,j}}\hat{T}_{1\rightarrow 2}\ket{\phi_{1,i}} .
$
Taking the simplest case of background permittivity modulation in dispersionless medium as an example, when the permittivity jumps from $\varepsilon_1$ to $\varepsilon_2$, the electric field transmittance and reflectance can be expressed as:
\begin{equation}
	T_E=\frac{\alpha_e}{2}+\frac 12\sqrt{\frac{\varepsilon_1}{\varepsilon_2}},\;R_E=\frac{\alpha_e}{2}-\frac 12\sqrt{\frac{\varepsilon_1}{\varepsilon_2}}.
\end{equation}
Here, by normalizing the basis vectors to a unit electric-field component, we obtain the scattering coefficients for the electric field. Assuming continuity of the electric displacement field yields $\alpha_e=\varepsilon_1/\varepsilon_2$, whereas assuming continuity of the electric field gives $\alpha_e =1$. It is observed that $R_E$ persists in this scenario.
In fact, adopting a mixed continuity condition between them enables us to choose the scaling factor $\alpha_e=\sqrt{\varepsilon_1/\varepsilon_2}$ to eliminate time reflection.

\begin{figure*}[htb]
	
	\subfigure{\includegraphics[width=0.32\linewidth]{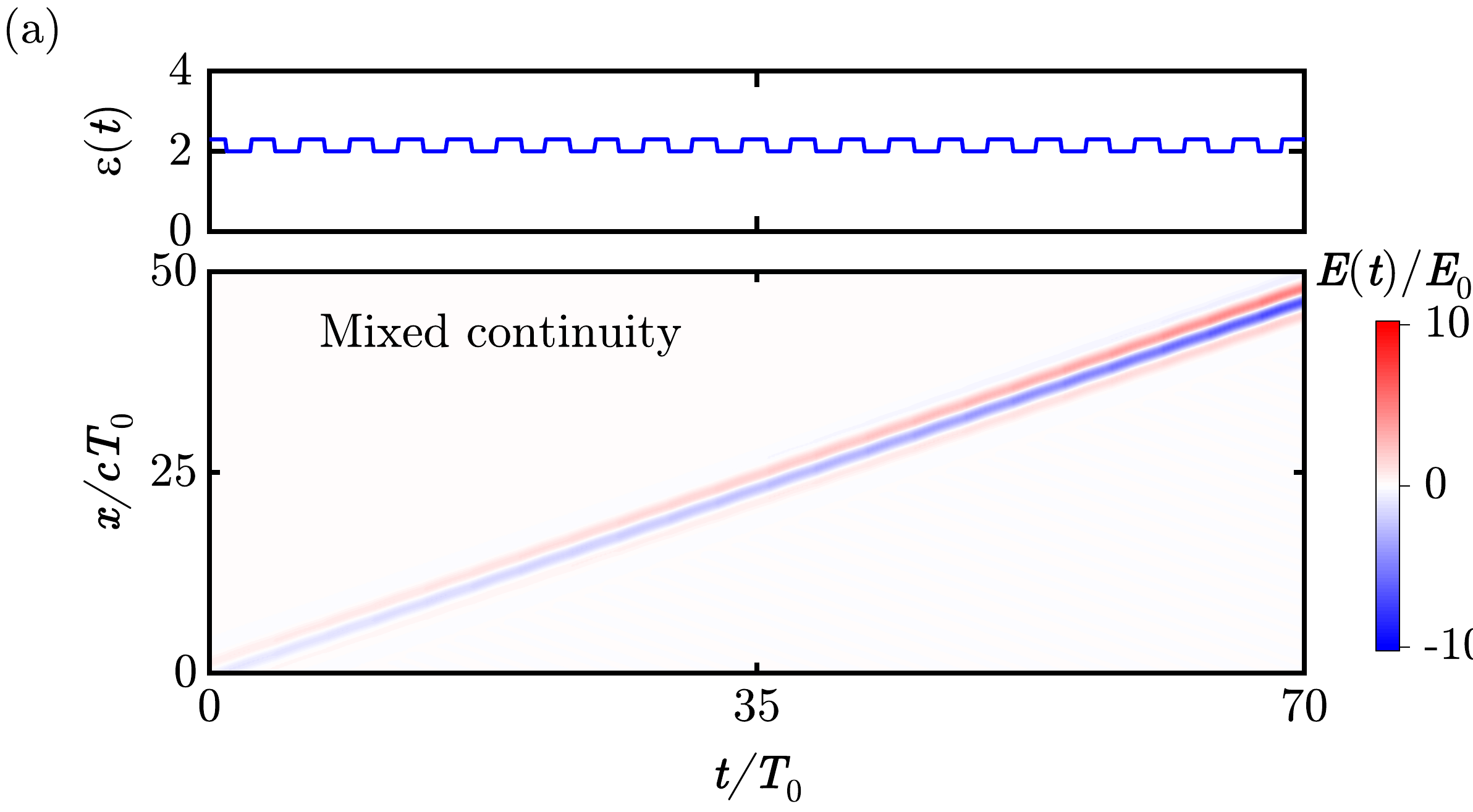}\label{f4a}}
	\subfigure{\includegraphics[width=0.32\linewidth]{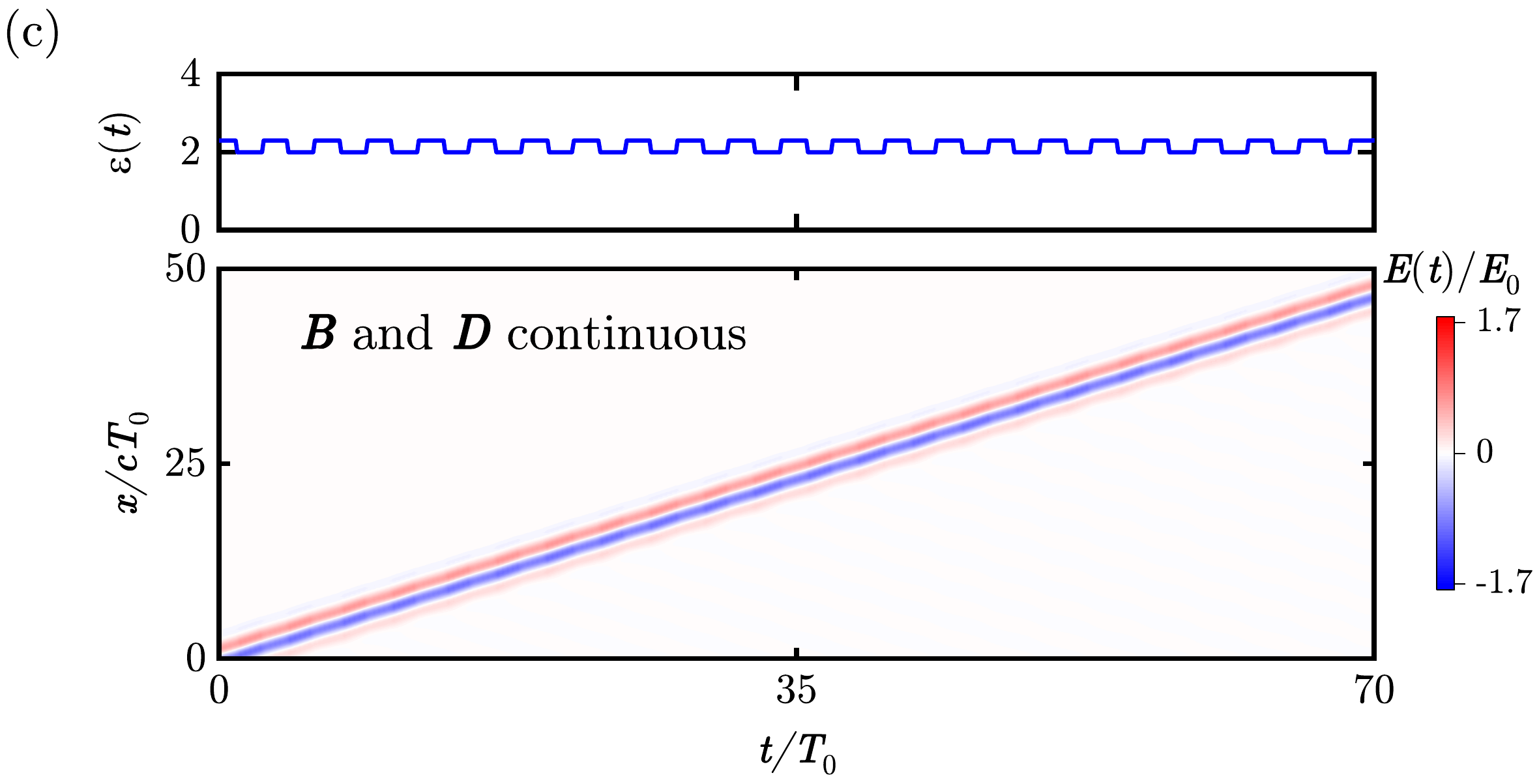}\label{f4c}}
	\subfigure{\includegraphics[width=0.32\linewidth]{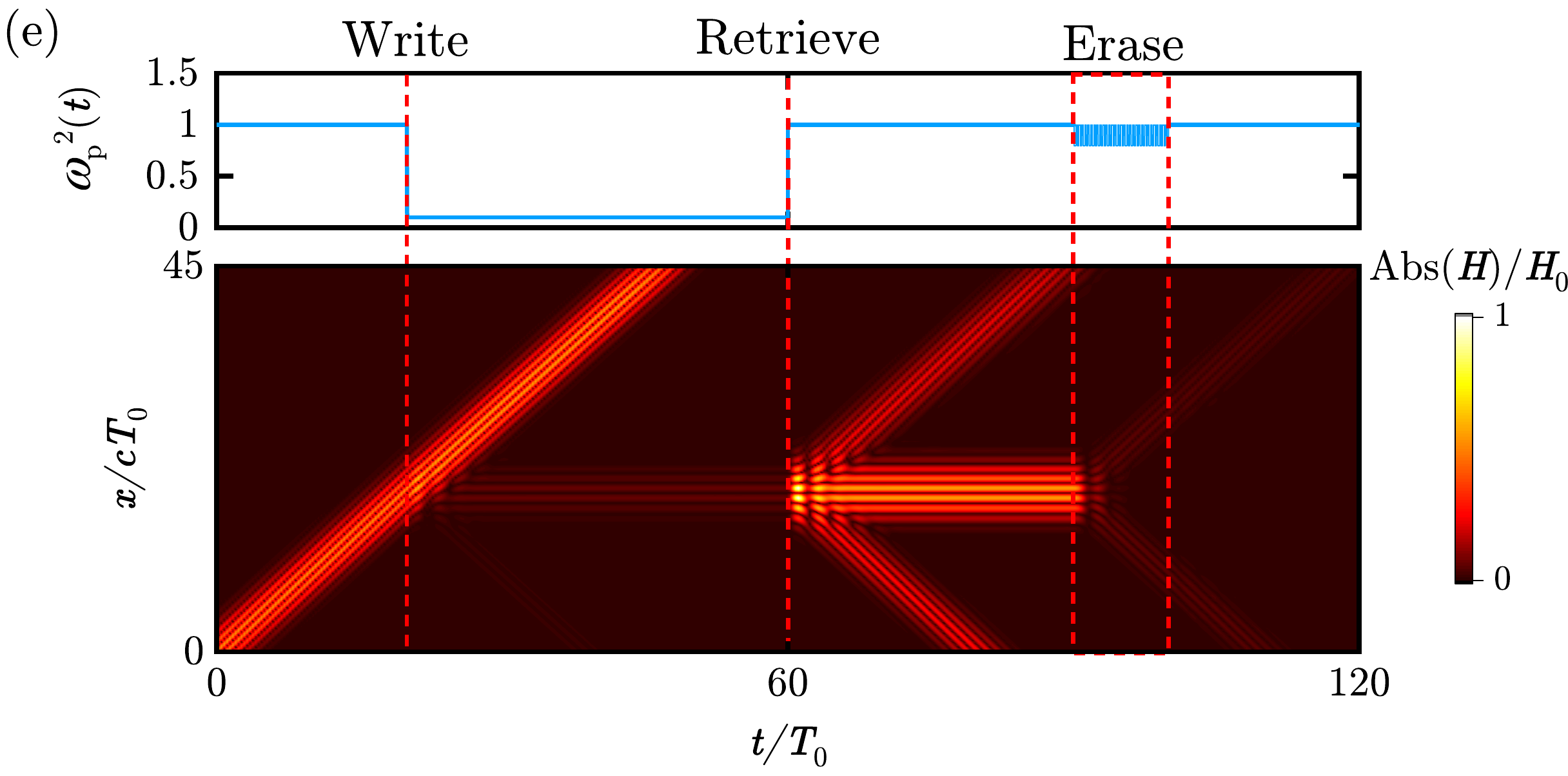}\label{f4e}}\\
	\subfigure{\includegraphics[width=0.32\linewidth]{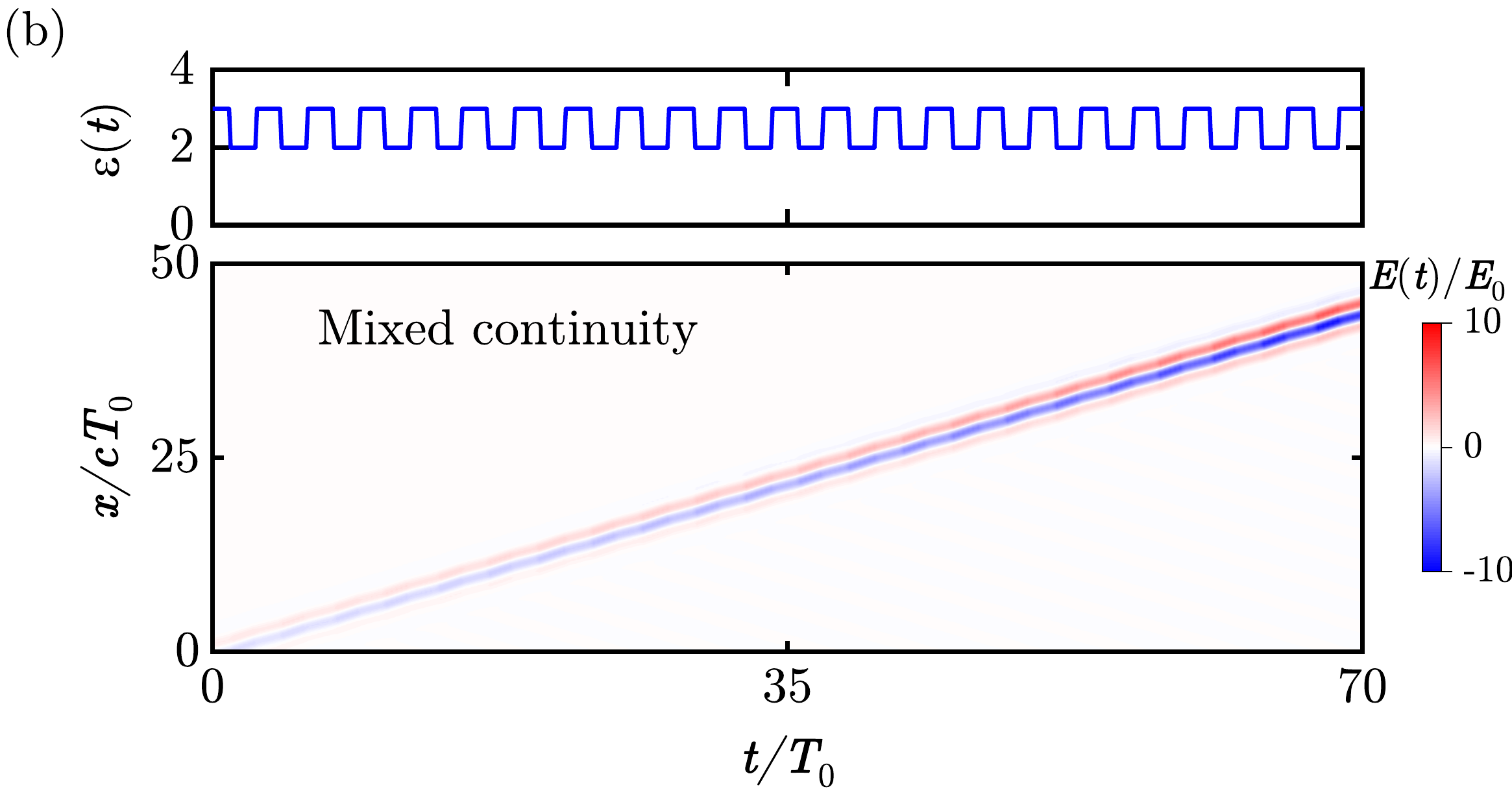}\label{f4b}}
	\subfigure{\includegraphics[width=0.32\linewidth]{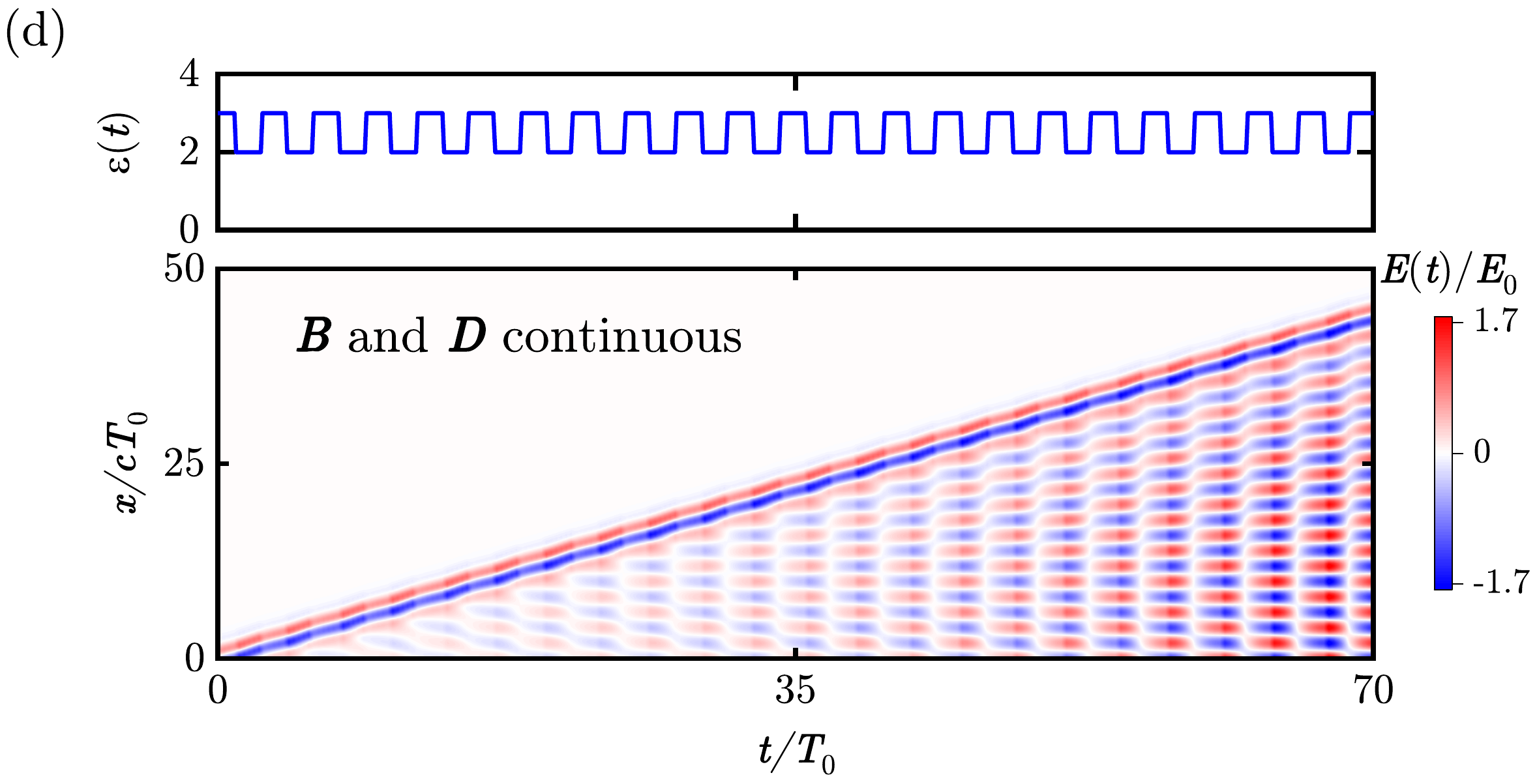}\label{f4d}}
	\subfigure{\includegraphics[width=0.32\linewidth]{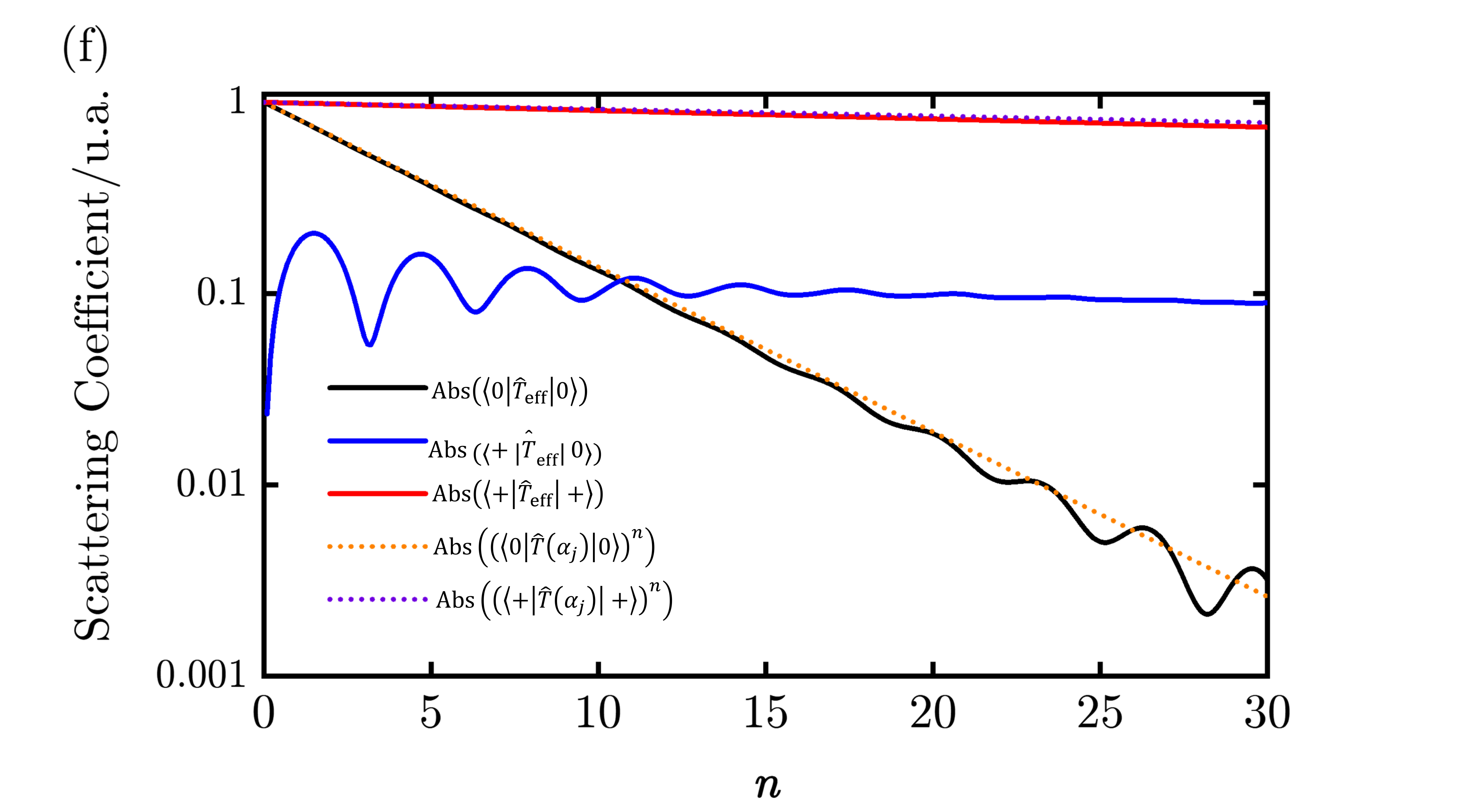}\label{f4f}}
	
	\caption{\label{fig4}Simulations showing pulse spatiotemporal evolution under different continuity conditions. (a)-(d) Spatiotemporal evolution of the electric field under simultaneous modulation of permittivity $\varepsilon$ and permeability $\mu$. In all cases, $\mu_1=1.1$, $\mu_2=1$. (a) and (c) correspond to $\varepsilon_1=2.3$, $\varepsilon_2=2 $ (b) and (d) correspond to increased permittivity modulation with $\varepsilon_1=3$, $\varepsilon_2=2 $.   Mixed continuity conditions are applied in (a) and (b), while (c) and (d) assume continuity of $\bm D$ and $\bm B$. (e) Pulse storage and retrieval enabled by temporal modulation of the plasma frequency. (f) Evolution of the matrix elements describing the scattering process with the number of modulation cycles.}
\end{figure*}

However, although setting $\alpha_e=\sqrt{\varepsilon_1/\varepsilon_2}$ eliminates temporal reflection, the case $R_E=0$ leads to $T_E=\sqrt{\varepsilon_1/\varepsilon_2}$. This is exactly the case of $\varepsilon^{0.5}E=\text{const}.$, which represents an adiabatic modulation. This indicates that under periodic modulation of permittivity, the traveling-wave fails to achieve net amplification. This limitation, however, can be overcome by introducing modulation of the permeability $\mu$. The corresponding transmission and reflection coefficients for the electric field can be expressed as
\begin{equation}
	T_E=\frac{\alpha_e}{2}+\frac {\alpha_m}{2}\sqrt{\frac{\varepsilon_1\mu_2}{\varepsilon_2\mu_1}},\;R_E=\frac{\alpha_e}{2}-\frac {\alpha_m}{2}\sqrt{\frac{\varepsilon_1\mu_2}{\varepsilon_2\mu_1}}.
\end{equation}
With $R_E=0$, it follows that $T_E=\alpha_e$ and thus $T_{E,1\rightarrow2}T_{E,2\rightarrow1}=\alpha_{e,1}\alpha_{e,2}$. The condition for this product to be greater than unity requires $\alpha_{e,2}^{-1}\neq \alpha_{e,1}$. The subscripts on $T_{E,1\rightarrow 2}$ and $\alpha_{e,1}$ denote quantities associated with a change from medium 1 to 2, and vice verse. Hence, we continue to assume tunability of the scaling factors. Specifically, we constrain the electric-field scaling factor to fall between the continuity of $\bm E$ and that of $\bm D $, and the magnetic-field scaling factor between the continuity of $\bm H$ and that of magnetic flux density $\bm D$. We further assume that $\varepsilon_1>\varepsilon_2$ and  $\mu_1>\mu_2$, which leads to the following constraints $\frac{\varepsilon_1}{\varepsilon_2}\geq\alpha_{e,1}\geq1$,  $\frac{\mu_1}{\mu_2}\geq\alpha_{m,1}\geq1$,  $1\geq\alpha_{e,2}\geq\frac{\varepsilon_2}{\varepsilon_1}$ and 
$1\geq\alpha_{m,2}\geq\frac{\mu_2}{\mu_1}$.

By further assuming that $\frac{\varepsilon_1}{\varepsilon_2}>\frac{\mu_1}{\mu_2}$, we can obtain the condition for maximizing $T_{E,1\rightarrow2}T_{E,2\rightarrow1}$: 
$
		\alpha_{m,1}=\frac{\mu_1}{\mu_2},\;\alpha_{e,1}=\sqrt{\frac{\varepsilon_1\mu_1}{\varepsilon_2\mu_2}},
		\alpha_{m,2}=1,\;\alpha_{e,2}=\sqrt{\frac{\varepsilon_2\mu_1}{\varepsilon_1\mu_2}}.
$
Thus, we finally arrive at the maximum transmission coefficient over one modulation cycle:
$
	T_{E,1\rightarrow2}T_{E,2\rightarrow1}=\alpha_{e,1}\alpha_{e,2}=\frac{\mu_1}{\mu_2}.
$

It is interesting to note that the mode growth efficiency depends exclusively on the permeability modulation ratio $\mu_1/\mu_2$, with the requirement for the permittivity modulation ratio being only implicitly present. In Fig.~\ref{fig4}, we compute the spatiotemporal evolution of the electric field under simultaneous modulation of permittivity and permeability. The calculation is performed using finite-difference time-domain (FDTD) methods.
 The source has a Gaussian profile, and the modulation frequency is chosen to be twice the central frequency of the source ($2\pi/T_0$). In Figs.~\ref{fig4}(a) and \ref{fig4}(c), we set $\varepsilon_1=2.3$, $\varepsilon_2=2$, $\mu_1=1.1$, and $\mu_2=1$. Fig.~\ref{fig4}(a) adopts the hybrid boundary condition, whereas Fig.~\ref{fig4}(c) assumes continuity of $D$ and $B$. It can be seen that Fig.~\ref{fig4}(a) exhibits reflectionless amplification of traveling waves, while Fig.~\ref{fig4}(c) shows no energy accumulation. In Figs.~\ref{fig4}(b) and \ref{fig4}(d), the permittivity modulation amplitude is increased by setting $\varepsilon_1=3$. The hybrid condition is applied in Fig.~\ref{fig4}(b), whereas Fig.~\ref{fig4}(d) shares the same continuity condition as Fig.~\ref{fig4}(c). Fig.~\ref{fig4}(b)shows reflectionless amplification of traveling waves with a growth rate identical to that in Fig.~\ref{fig4}(a). This confirms that the amplification rate of traveling waves depends only on the ratio $\mu_1/\mu_2$. For Fig.~\ref{fig4}(d), the larger modulation amplitude gives rise to a momentum band gap, leading to the amplification of standing waves. The amplification mechanism enabled by the hybrid continuity condition is therefore non-resonant, broadband, and reflectionless, fundamentally distinct from that in photonic time crystals. Thus greatly expanding the potential of temporal modulation for signal amplification. Conversely, the time-reversed counterpart of this process enables reflectionless wave absorption.

In Drude media, the mixed continuity condition can also lift the limitations in temporal modulation. We label the three eigenstates in medium $i$ corresponding to $\omega=0$, $\omega_-$, and $\omega_+$ as $\ket{\phi_{i,0}}$, $\ket{\phi_{i,-}}$, and $\ket{\phi_{i,+}}$, respectively. The non-orthogonality of the eigenmodes leads to a non-reciprocal relation between the inner products$
		\braket{\phi^{1,i}}{\phi_{2,j}}\neq \braket{\phi^{2,j}}{\phi_{1,i}}.
$
Thus, although it is well known that in Drude media a temporal parameter jump can convert traveling waves into static magnetic fields.  This can be evidenced by the fact that $\braket{\phi^{2,0}}{\phi_{1,\pm}}$ is nonzero. However, under plasma frequency modulation, the zero-frequency (DC) mode remains independent of the dielectric parameters, namely, $
		\ket{\phi_{1,0}}=\ket{\phi_{2,0}}.$
Consequently, 
$
		\braket{\phi^{2,\pm}}{\phi_{1,0}}=\braket{\phi^{2,\pm}}{\phi_{2,0}}=0.
$
This suggests that the conversion from static magnetic fields to propagating waves is prohibited under current continuity conditions. This prohibition of mode conversion is analogous to a selection rule. Fortunately, this limitation can be overcome by modifying the condition of current continuity, i.e., making $\bra{\phi^{2,\pm}}\hat T\ket{\phi_{1,0}}$ nonzero. 
Specifically, the coefficients for the conversion from oscillatory to static fields and vice versa can be written as follows:
\begin{equation}
	\begin{aligned}
	\bra{\phi^{2,0}}\hat T(\alpha_j)\ket{\phi_{1,\pm}}=\frac{(\alpha_j -1)\omega_{p,2}^2}{2(c^2k_y^2+\omega_{p,2}^2)}		\label{eq31}
	\end{aligned}
\end{equation}
and
\begin{equation}
	\begin{aligned}
		\bra{\phi^{2,\pm}}\hat T(\alpha_j)\ket{\phi_{1,0}}=\frac{c^2k_y^2(\alpha_j\omega_{p,1}^2-\omega_{p,2}^2)}{(c^2k_y^2+\omega_{p,2}^2)\omega_{p,1}^2}.	\label{eq32}
	\end{aligned}
\end{equation}
Here, the basis vectors are normalized to a unit current density component. 
It is evident that the conversion from oscillatory to static fields is prohibited when $\bm J /\omega_{p}^2$ is continuous, i.e., $\alpha_j=\omega_{p,2}^2/\omega_{p,1}^2$. This conversion thus provides a mechanism for optical pulse storage, with potential applications in optical computing.

We now provide a detailed description of this scheme. The spatiotemporal evolution of the pulse is presented in Fig.~\ref{fig4}(e). Here, the initial plasma frequency is set to unity ($\Omega_0=2\pi/T_0 $), and a Gaussian pulse with a center frequency of $3\Omega_0$ is adopted to minimize dispersive effects. The writing process involves a decrease in plasma density to $\omega_{p,2}^2/\omega_{p,1}^2=0.1$, with current density kept continuous. The retrieval process, as illustrated in Fig.~\ref{fig4}(e), restores the plasma frequency to its initial value while keeping $\bm J/\omega_{p}^2$ conserved, i.e., $\alpha_j=\omega_{p,1}^2/\omega_{p,2}^2$. By comparing Eq.~\ref{eq31} and Eq.~\ref{eq32}, it can be seen that the retrieved waveform is, in fact, an approximation of the second spatial derivative of the original pulse. Notably, the presented storage mechanism does not freeze the pulse profile itself, but rather encodes its second-order derivative into static fields. This feature offers a native interface for differential photonic computing.

To erase the signal, we employ a multi-cycle, high-frequency modulation of the plasma density, described by:
\begin{equation}
	\begin{aligned}
		&\bra{\phi^{a,i}}\hat T_\text{eff}\ket{\phi_{a,\pm}}\\
		&=\bra{\phi^{a,i}}(\hat T_\text{step})^n\ket{\phi_{a,\pm}}\\
		&=\bra{\phi^{a,i}}(\mathrm e^{-\mathrm i\hat {M}_a\Delta t}\hat{T}_{b\rightarrow a}\mathrm e^{-\mathrm i\hat {M}_b\Delta t}\hat T_{a\rightarrow b})^n\ket{\phi_{a,\pm}}\label{eq33}.
	\end{aligned}
\end{equation}
Here, $\hat {T}_\text{eff}$ characterizes the equivalent scattering process. $n$ is number of modulation cycles. Under the assumption $\omega_{p,a}>\omega_{p,b}$, the modulation scheme involves periodic alternation between $\omega_{p,a}$ and $\omega_{p,b}$, with each state lasting for a duration $\Delta t$. For the downward transition $\omega_{p,a}\rightarrow\omega_{p,b}$ (corresponding to a decrease in plasma density), we set $\hat{T}_{a\rightarrow b}=\hat{T}(\alpha_j)$, where $\alpha_j=\omega_{p,b}^2/\omega_{p,a}^2$. For the reverse transition (increase in plasma density), the current is taken to be continuous, so that the transfer matrix 
$\hat{T}_{b\rightarrow a}=\hat I$ is the identity transformation.

Further, since $\hat{T}_\text{step}$ is diagonalizable, there exists an invertible matrix $P$ such that $\hat{T}_\text{step} = P \Lambda P^{-1}$, where $\Lambda=\mathrm {diag}(\lambda_0,\lambda_-,\lambda_+)$ is the diagonal matrix with the corresponding eigenvalues $\lambda_i$ on its diagonal.
Then we have $
\hat T_\text{eff}=P\Lambda^n P^{-1}.
$

When the modulation frequency is high (i.e.,$\omega_\pm\Delta t\ll 1$) and the modulation amplitude is small ($\omega_{p,a}\simeq\omega_{p,b}$), the diagonal elements of $P$ approach unity, and the eigenvalues $\lambda_i$ can be approximated by $\lambda_i \simeq \bra{\phi^i}\hat{T}(\alpha_j)\ket{\phi_i}$. Fig.~\ref{fig4}(f) shows these matrix elements as functions of the cycle number $n$, comparing the exact results (solid lines) with the analytical approximation (dotted lines). This comparison demonstrates that the approximation successfully captures the amplitude dynamics of the modes.

Specifically, for large values of $k_y$, the decay of static fields is much faster than the propagating wave (see Supplementary Material). This verifies that, even far away from the momentum band gap, mode separation still occurs under such mixed continuity conditions as aforementioned, enabling mode selection. As shown in Fig.~\ref{fig4}(e), after the periodic modulation described above, the static field decays exponentially and is thus erased.

In conclusion, this work has developed a comprehensive framework for time-varying dispersive media. By transcending the conventional assumption of fixed field continuity, we introduce a new degree of freedom that alters the system's response. We have shown that this additional freedom enables phenomena previously thought impossible under pure temporal modulation. Notably, we demonstrate broadband, reflectionless amplification of traveling waves, which fundamentally differs from photonic time crystals that rely on momentum bandgaps. Furthermore, we achieve controlled conversion between static and propagating fields, a functionality with direct implications for optical pulse storage and processing. Both effects are shown to be unattainable under conventional single-continuity conditions with purely temporal modulation. These findings not only enrich time-varying electromagnetics but also open promising avenues for on-chip amplification, optical computing, and communications.
\vspace{1em}
\begin{acknowledgments}
	\textit{Acknowledgments}---This work is supported by the National Key R\&D Program of China (No.2022YFE0204100), National Natural Science Foundation of China (NSFC, Contracts No.  12105251, No. 12175050 and 12205067).
\end{acknowledgments}
\vspace{1em}
\begin{acknowledgments}
\textit{Data availability}---The data that support the findings of this study are available from the corresponding author upon reasonable request.
\end{acknowledgments}

\bibliography{ref}

@article{PhysRevLett.132.263802,
	title = {Time Refraction and Time Reflection above Critical Angle for Total Internal Reflection},
	author = {Bar-Hillel, Lior and Dikopoltsev, Alex and Kam, Amit and Sharabi, Yonatan and Segal, Ohad and Lustig, Eran and Segev, Mordechai},
	journal = {Phys. Rev. Lett.},
	volume = {132},
	issue = {26},
	pages = {263802},
	numpages = {6},
	year = {2024},
	month = {Jun},
	publisher = {American Physical Society},
	doi = {10.1103/PhysRevLett.132.263802},
	url = {https://link.aps.org/doi/10.1103/PhysRevLett.132.263802}
}

@article{PhysRevLett.133.186902,
	title = {Harnessing the Natural Resonances of Time-Varying Dispersive Interfaces},
	author = {Rizza, Carlo and Vincenti, Maria Antonietta and Castaldi, Giuseppe and Contestabile, Alessandra and Galdi, Vincenzo and Scalora, Michael},
	journal = {Phys. Rev. Lett.},
	volume = {133},
	issue = {18},
	pages = {186902},
	numpages = {7},
	year = {2024},
	month = {Oct},
	publisher = {American Physical Society},
	doi = {10.1103/PhysRevLett.133.186902},
	url = {https://link.aps.org/doi/10.1103/PhysRevLett.133.186902}
}

@article{ WOS:001287600900019,
	Author = {Jones, Thomas R. and Kildishev, Alexander V. and Segev, Mordechai and
	Peroulis, Dimitrios},
	Title = {Time-reflection of microwaves by a fast optically-controlled
	time-boundary},
	Journal = {Nat. Commun.},
	Year = {2024},
	Volume = {15},
	Number = {1},
	pages={6786},
	Month = {AUG 8},
	doi = {10.1038/s41467-024-51171-6},
	url={doi.org/10.1038/s41467-024-51171-6}
}

@article{ WOS:001604755900027,
	Author = {Wang, Shaoyun and Shao, Nan and Chen, Hui and Chen, Jiaji and Qian,
	Honghua and Wu, Qian and Duan, Huiling and Alu, Andrea and Huang,
	Guoliang},
	Title = {Experimental realization of temporal refraction and reflection in
	elastic beams},
	Journal = {Nat. Commun.},
	Year = {2025},
	Volume = {16},
	Number = {1},
	pages={9520 },
	Month = {OCT 28},
	DOI = {10.1038/s41467-025-64530-8},}

@article{ WOS:001575855000001,
	Author = {Galiffi, Emanuele and Solis, Diego Martinez and Yin, Shixiong and
	Engheta, Nader and Alu, Andrea},
	Title = {Electrodynamics of photonic temporal interfaces},
	Journal = {LIGHT-SCIENCE \& APPLICATIONS},
	Year = {2025},
	Volume = {14},
	Number = {1},
	Month = {SEP 23},
	DOI = {10.1038/s41377-025-01947-2}}

@article{
	doi:10.1126/sciadv.adz5445,
	author = {Yanyan He  and Zhaohui Dong  and Guangzhen Li  and Penghong Yu  and Xiaoxiong Wu  and Xianfeng Chen  and Luqi Yuan },
	title = {Observing momentum conservation at temporal interfaces in synthetic frequency dimension},
	journal = {Science Advances},
	volume = {11},
	number = {51},
	pages = {eadz5445},
	year = {2025},
	doi = {10.1126/sciadv.adz5445},
	URL = {https://www.science.org/doi/abs/10.1126/sciadv.adz5445},}

@article{PhysRevLett.125.127403,
	title = {Wood Anomalies and Surface-Wave Excitation with a Time Grating},
	author = {Galiffi, Emanuele and Wang, Yao-Ting and Lim, Zhen and Pendry, J. B. and Al\`u, Andrea and Huidobro, Paloma A.},
	journal = {Phys. Rev. Lett.},
	volume = {125},
	issue = {12},
	pages = {127403},
	numpages = {6},
	year = {2020},
	month = {Sep},
	publisher = {American Physical Society},
	doi = {10.1103/PhysRevLett.125.127403},
	url = {https://link.aps.org/doi/10.1103/PhysRevLett.125.127403}
}

@article{
	doi:10.1073/pnas.2119705119,
	author = {Alex Dikopoltsev  and Yonatan Sharabi  and Mark Lyubarov  and Yaakov Lumer  and Shai Tsesses  and Eran Lustig  and Ido Kaminer  and Mordechai Segev },
	title = {Light emission by free electrons in photonic time-crystals},
	journal = {Proceedings of the National Academy of Sciences},
	volume = {119},
	number = {6},
	pages = {e2119705119},
	year = {2022},
	doi = {10.1073/pnas.2119705119},
	URL = {https://www.pnas.org/doi/abs/10.1073/pnas.2119705119},}

@article{ WOS:000531425900022,
	Author = {Zhou, Yiyu and Alam, M. Zahirul and Karimi, Mohammad and Upham, Jeremy
	and Reshef, Orad and Liu, Cong and Willner, Alan E. and Boyd, Robert W.},
	Title = {Broadband frequency translation through time refraction in an
	epsilon-near-zero material},
	Journal = {Nat. Commun.},
	Year = {2020},
	Volume = {11},
	pages={2180 },
	Number = {1},
	Month = {MAY 1},
	DOI = {10.1038/s41467-020-15682-2},
	Article-Number = {2180},}

@article{ WOS:001258302000023,
	Author = {Qin, Chengzhi and Ye, Han and Wang, Shulin and Zhao, Lange and Liu,
	Menglin and Li, Yinglan and Hu, Xinyuan and Liu, Chenyu and Wang, Bing
	and Longhi, Stefano and Lu, Peixiang},
	Title = {Observation of discrete-light temporal refraction by moving potentials
	with broken Galilean invariance},
	Journal = {Nat. Commun.},
	Year = {2024},
	pages={5444},
	Volume = {15},
	Number = {1},
	Month = {JUN 27},
	DOI = {10.1038/s41467-024-49747-3},
	Article-Number = {5444},
	EISSN = {2041-1723},}

@article{PhysRevLett.126.163902,
	title = {Disordered Photonic Time Crystals},
	author = {Sharabi, Yonatan and Lustig, Eran and Segev, Mordechai},
	journal = {Phys. Rev. Lett.},
	volume = {126},
	issue = {16},
	pages = {163902},
	numpages = {6},
	year = {2021},
	month = {Apr},
	publisher = {American Physical Society},
	doi = {10.1103/PhysRevLett.126.163902},
	url = {https://link.aps.org/doi/10.1103/PhysRevLett.126.163902}
}

@article{PhysRevLett.128.094503,
	title = {Experimental Implementation of Wave Propagation in Disordered Time-Varying Media},
	author = {Apffel, Benjamin and Wildeman, Sander and Eddi, Antonin and Fort, Emmanuel},
	journal = {Phys. Rev. Lett.},
	volume = {128},
	issue = {9},
	pages = {094503},
	numpages = {7},
	year = {2022},
	month = {Mar},
	publisher = {American Physical Society},
	doi = {10.1103/PhysRevLett.128.094503},
	url = {https://link.aps.org/doi/10.1103/PhysRevLett.128.094503}
}

@article{
	doi:10.1126/science.abo3324,
	author = {Mark Lyubarov  and Yaakov Lumer  and Alex Dikopoltsev  and Eran Lustig  and Yonatan Sharabi  and Mordechai Segev },
	title = {Amplified emission and lasing in photonic time crystals},
	journal = {Science},
	volume = {377},
	number = {6604},
	pages = {425-428},
	year = {2022},
	doi = {10.1126/science.abo3324},
}

@article{10.1063/1.4928659,
	author = {Reyes-Ayona, J. R. and Halevi, P.},
	title = {Observation of genuine wave vector (k or $\beta$) gap in a dynamic transmission line and temporal photonic crystals},
	journal = {Applied Physics Letters},
	volume = {107},
	number = {7},
	pages = {074101},
	year = {2015},
	month = {08},
	doi = {10.1063/1.4928659},
	url = {https://doi.org/10.1063/1.4928659},
}

@article{PhysRevA.79.053821,
	title = {Reflection and transmission of a wave incident on a slab with a time-periodic dielectric function $\varepsilon (t)$},
	author = {Zurita-S\'anchez, Jorge R. and Halevi, P. and Cervantes-Gonz\'alez, Juan C.},
	journal = {Phys. Rev. A},
	volume = {79},
	issue = {5},
	pages = {053821},
	numpages = {13},
	year = {2009},
	month = {May},
	publisher = {American Physical Society},
	doi = {10.1103/PhysRevA.79.053821},
	url = {https://link.aps.org/doi/10.1103/PhysRevA.79.053821}
}

@article{Lustig:18,
	author = {Eran Lustig and Yonatan Sharabi and Mordechai Segev},
	journal = {Optica},
	number = {11},
	pages = {1390--1395},
	publisher = {Optica Publishing Group},
	title = {Topological aspects of photonic time crystals},
	volume = {5},
	month = {Nov},
	year = {2018},
	url = {https://opg.optica.org/optica/abstract.cfm?URI=optica-5-11-1390},
	doi = {10.1364/OPTICA.5.001390}}

@article{
	doi:10.1126/sciadv.adg7541,
	author = {Xuchen Wang  and Mohammad Sajjad Mirmoosa  and Viktar S. Asadchy  and Carsten Rockstuhl  and Shanhui Fan  and Sergei A. Tretyakov },
	title = {Metasurface-based realization of photonic time crystals},
	journal = {Science Advances},
	volume = {9},
	number = {14},
	pages = {eadg7541},
	year = {2023},
	doi = {10.1126/sciadv.adg7541},
	URL = {https://www.science.org/doi/abs/10.1126/sciadv.adg7541},}

@article{PhysRevB.103.144303,
	title = {Functional analysis of the polarization response in linear time-varying media: A generalization of the Kramers-Kronig relations},
	author = {Sol\'{\i}s, Diego M. and Engheta, Nader},
	journal = {Phys. Rev. B},
	volume = {103},
	issue = {14},
	pages = {144303},
	numpages = {14},
	year = {2021},
	month = {Apr},
	publisher = {American Physical Society},
	doi = {10.1103/PhysRevB.103.144303},
	url = {https://link.aps.org/doi/10.1103/PhysRevB.103.144303}
}

@article{PhysRevLett.134.183801,
	title = {Plasmonic Time Crystals},
	author = {Feinberg, Joshua and Fernandes, David E. and Shapiro, Boris and Silveirinha, M\'ario G.},
	journal = {Phys. Rev. Lett.},
	volume = {134},
	issue = {18},
	pages = {183801},
	numpages = {7},
	year = {2025},
	month = {May},
	publisher = {American Physical Society},
	doi = {10.1103/PhysRevLett.134.183801},
	url = {https://link.aps.org/doi/10.1103/PhysRevLett.134.183801}
}

@article{li2023stationary,
	title={Stationary charge radiation in anisotropic photonic time crystals},
	author={Li, Huanan and Yin, Shixiong and He, Huan and Xu, Jingjun and Al{\`u}, Andrea and Shapiro, Boris},
	journal={Physical Review Letters},
	volume={130},
	number={9},
	pages={093803},
	year={2023},
	publisher={APS}
}

@article{PhysRevLett.128.173901,
	title = {Nonreciprocity and Faraday Rotation at Time Interfaces},
	author = {Li, Huanan and Yin, Shixiong and Al\`u, Andrea},
	journal = {Phys. Rev. Lett.},
	volume = {128},
	issue = {17},
	pages = {173901},
	numpages = {7},
	year = {2022},
	month = {Apr},
	publisher = {American Physical Society},
	doi = {10.1103/PhysRevLett.128.173901},
}

@article{he2023faraday,
	title={Faraday rotation in nonreciprocal photonic time-crystals},
	author={He, Huan and Zhang, Sihao and Qi, Jiwei and Bo, Fang and Li, Huanan},
	journal={Applied Physics Letters},
	volume={122},
	number={5},
	year={2023},
	publisher={AIP Publishing}
}

@article{mirmoosa2022dipole,
	title={Dipole polarizability of time-varying particles},
	author={Mirmoosa, Mohammad Sajjad and Koutserimpas, TT and Ptitcyn, GA and Tretyakov, SA and Fleury, R},
	journal={New Journal of Physics},
	volume={24},
	number={6},
	pages={063004},
	year={2022},
	publisher={IOP Publishing}
}

@article{Sharabi:22,
	author = {Yonatan Sharabi and Alex Dikopoltsev and Eran Lustig and Yaakov Lumer and Mordechai Segev},
	journal = {Optica},
	number = {6},
	pages = {585--592},
	publisher = {Optica Publishing Group},
	title = {Spatiotemporal photonic crystals},
	volume = {9},
	month = {Jun},
	year = {2022},
	url = {https://opg.optica.org/optica/abstract.cfm?URI=optica-9-6-585},
	doi = {10.1364/OPTICA.455672},
}

@article{
	doi:10.1126/science.aah6822,
	author = {Hanan Herzig Sheinfux  and Yaakov Lumer  and Guy Ankonina  and Azriel Z. Genack  and Guy Bartal  and Mordechai Segev },
	title = {Observation of Anderson localization in disordered nanophotonic structures},
	journal = {Science},
	volume = {356},
	number = {6341},
	pages = {953-956},
	year = {2017},
	doi = {10.1126/science.aah6822},
	URL = {https://www.science.org/doi/abs/10.1126/science.aah6822}}

@article{GaliffiYinAl,
	url = {https://doi.org/10.1515/nanoph-2022-0200},
	title = {Tapered photonic switching},
	title = {},
	author = {Emanuele Galiffi and Shixiong Yin and Andrea Alú},
	pages = {3575--3581},
	volume = {11},
	number = {16},
	journal = {Nanophotonics},
	doi = {doi:10.1515/nanoph-2022-0200},
	year = {2022},
	lastchecked = {2026-03-10}
}

@article{10.1119/1.1970365,
	author = {Whiteside, Haven},
	title = {Newton's Derivation of the Velocity of Sound},
	journal = {American Journal of Physics},
	volume = {32},
	number = {5},
	pages = {384-384},
	year = {1964},
	month = {05},
	issn = {0002-9505},
	doi = {10.1119/1.1970365},
	url = {https://doi.org/10.1119/1.1970365}
}

@book{pierce2019acoustics,
	title={Acoustics: an introduction to its physical principles and applications},
	author={Pierce, Allan D},
	year={2019},
	publisher={Springer}
}

@book{lighthill2001waves,
	title={Waves in fluids},
	author={Lighthill, Michael J and Lighthill, James},
	year={2001},
	publisher={Cambridge university press}
}

@book{newton1833philosophiae,
	title={Philosophiae naturalis principia mathematica},
	author={Newton, Isaac},
	volume={1},
	year={1833},
	publisher={G. Brookman}
}

@article{laplace1816vitesse,
	title={Sur la vitesse du son dans l’air et dans l’eau},
	author={Laplace, PS de},
	journal={Ann. Chim. Phys},
	volume={3},
	number={2},
	pages={238--241},
	year={1816}
}
\end{document}